\documentclass[sigconf]{acmart}
\usepackage{soul}
\usepackage{hyperref}
\usepackage{enumitem}
\usepackage{listings}
\usepackage{cleveref}
\usepackage{booktabs} 
\usepackage{xspace}
\usepackage{xcolor}
\usepackage{microtype}
\usepackage{subcaption}
\usepackage{balance}
\usepackage{mathtools}
\usepackage{scalerel}
\usepackage[makeroom]{cancel}

\usepackage{tcolorbox}
\usepackage{microtype}
\usepackage[section]{placeins}
\usepackage{graphicx}
\usepackage[T1]{fontenc}
\usepackage{svg}
\usepackage{amsmath}
\usepackage{amsmath}
\usepackage[mathletters]{ucs}
\usepackage[T1]{fontenc}
\usepackage[utf8x]{inputenc}
\usepackage{pict2e}
\usepackage{wasysym}
\usepackage[english]{babel}
\usepackage{tikz}
\usepackage{array}

\usepackage{algpseudocode}
\usepackage{tablefootnote}
\usepackage[normalem]{ulem}

\DeclareMathSymbol{:}{\mathord}{operators}{"3A}

\usepackage{multirow}
 \usepackage{balance}

\def\HS{\hspace{\fontdimen2\font}}
\definecolor{darkgreen}{rgb}{0.15,0.55,0.15}
\definecolor{darkblue}{rgb}{0.1,0.1,0.5}
\definecolor{blue}{rgb}{0.01,0.40,.8}
\definecolor{darkgreen}{rgb}{0.15,0.55,0.15}
\definecolor{mred}{rgb}{.80,.12,.30}
\definecolor{grey}{rgb}{0.5,0.5,0.5}
\definecolor{Purple}{rgb}{.75,0,.85}
\definecolor{light-gray}{gray}{0.95}
\definecolor{mid-gray}{gray}{0.85}
\definecolor{darkred}{rgb}{0.7,0.25,0.25}
\definecolor{rose}{rgb}{1.0, 0.01, 0.24}
\newcommand{\red}[1]{\textcolor{red}{#1}}

\newcommand{\blue}[1]{\textcolor{blue}{#1}}
\usepackage{algorithm}

\newtcbox{\redbox}{on line,
  colframe=white,colback=red!10!white,
  height=1em,valign=bottom,
  boxrule=0.5pt,arc=2pt,boxsep=0pt,left=2pt,right=2pt,top=1pt,bottom=1pt}
\newtcbox{\bluebox}{on line,
  colframe=white,colback=blue!10!white,
  height=1em,valign=bottom,
  boxrule=0.5pt,arc=2pt,boxsep=0pt,left=2pt,right=2pt,top=1pt,bottom=1pt}

\newcommand{\eat}[1]{}

\newcommand{\stitle}[1]{\smallskip\noindent\textbf{#1}}

\newtheorem{prop}{Proposition}
\newtheorem*{proofk}{Proof Sketch}
\newtheorem{ex}{Example}

\newtheorem{example}[ex]{Example}

\newlength{\listingindent}                
\usepackage[LGRgreek]{mathastext}

\setlist{leftmargin=*}

\graphicspath{ {images/} }
\setcopyright{none} 

\renewcommand\footnotetextcopyrightpermission[1]{} 
\begin{document}

\newcommand{\total}[0]{\emph{TOTAL}\xspace}
\newcommand{\cnt}[0]{\emph{COUNT}\xspace}
\newcommand{\cof}[0]{\emph{COF}\xspace}
\newcommand{\cjt}[0]{\texttt{CJT}\xspace}
\newcommand{\cjts}[0]{\texttt{CJT}s\xspace}
\newcommand{\jt}[0]{\texttt{JT}\xspace}
\newcommand{\steiner}[0]{\texttt{ST}\xspace}
\newcommand{\jtivm}[0]{\texttt{JT-IVM}\xspace}
\newcommand{\jts}[0]{\texttt{JT}s\xspace}
\newcommand{\analytics}[0]{\emph{DONT USE}\xspace}
\newcommand{\dashq}[0]{\emph{dashboard query}\xspace}
\newcommand{\dashqs}[0]{\emph{dashboard queries}\xspace}
\newcommand{\intq}[0]{\emph{interaction query}\xspace}
\newcommand{\intqs}[0]{\emph{interaction queries}\xspace}
\newcommand{\AnnoP}[0]{${\bf A_p}$\xspace}
\newcommand{\Anno}[0]{${\bf A}$\xspace}
\newcommand{\AnnoDiff}[0]{${\bf A_\Delta}$\xspace}
\newcommand{\BagDiff}[0]{${\bf B_\Delta}$\xspace}

\newcommand{\ewu}[1]{\red{ewu: #1\xspace}}
\newcommand{\sys}[0]{\texttt{Treant}\xspace}
\newcommand{\matlab}[0]{\texttt{Matlab}\xspace}
\newcommand{\revise}[2]{{#1\xspace}}
\newcommand{\techreport}[1]{}

\newenvironment{myitemize}{%
\begin{itemize}[leftmargin=1em, itemsep=.1em, parsep=.1em, topsep=.1em,
    partopsep=.1em]}
{\end{itemize}}

\author{Zezhou Huang}
\affiliation{
  \institution{Columbia University}
}
\email{zh2408@columbia.edu}
\author{Eugene Wu}
\affiliation{
  \institution{DSI, Columbia University}
}
\email{ewu@cs.columbia.edu}

\begin{abstract}
Dashboards are vital in modern business intelligence tools, providing non-technical users with an interface to access comprehensive business data. With the rise of cloud technology, there is an increased number of data sources to provide enriched contexts for various analytical tasks, leading to a demand for interactive dashboards over a large number of joins.
Nevertheless, joins are among the most expensive operations in DBMSes, making the support of interactive dashboards over joins challenging.

In this paper, we present \sys, a dashboard accelerator for queries over large joins. 
\sys uses factorized query execution to handle aggregation queries over large joins, which alone is still insufficient for interactive speeds. 
To address this, we exploit the incremental nature of user interactions using Calibrated Junction Hypertree (\cjt), a novel data structure that applies lightweight materialization of the intermediates during  factorized execution. 
\cjt ensures that the work needed to compute a query is proportional to {\it how different it is from the previous query}, rather than the overall complexity.
\sys manages \cjts to share work between queries and performs materialization offline or during user "think-times." 
Implemented as a middleware that rewrites SQL, \sys is portable to any SQL-based DBMS. 
Our experiments on single node and cloud DBMSes show that \sys improves  dashboard interactions by two orders of magnitude, and provides $10{\times}$ improvement for ML augmentation compared to SOTA factorized ML system.

\end{abstract}

\title{Lightweight Materialization for Fast Dashboards Over Joins}

\maketitle
\pagestyle{plain}

\vspace{-3mm}
\section{Introduction}
\label{sec:intro}

\begin{figure}
  \centering
  \begin{subfigure}[t]{0.46\textwidth}
         \centering
         \includegraphics[width=\textwidth]{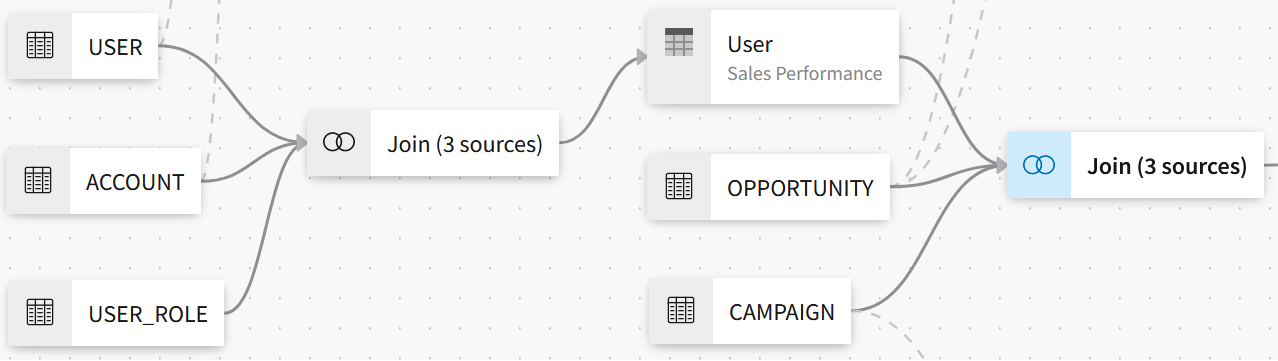}
         \vspace*{-5mm}
          \caption{Build join graph offline by data engineering.  Three tables related to \texttt{User} are first joined. Then, the enriched \texttt{User} table is joined with \texttt{Opportunity} and \texttt{Campaign}.}
          \label{fig:sigmajoin}
     \end{subfigure}
    \begin{subfigure}[t]{0.46\textwidth}
         \centering
         \includegraphics[width=\textwidth]{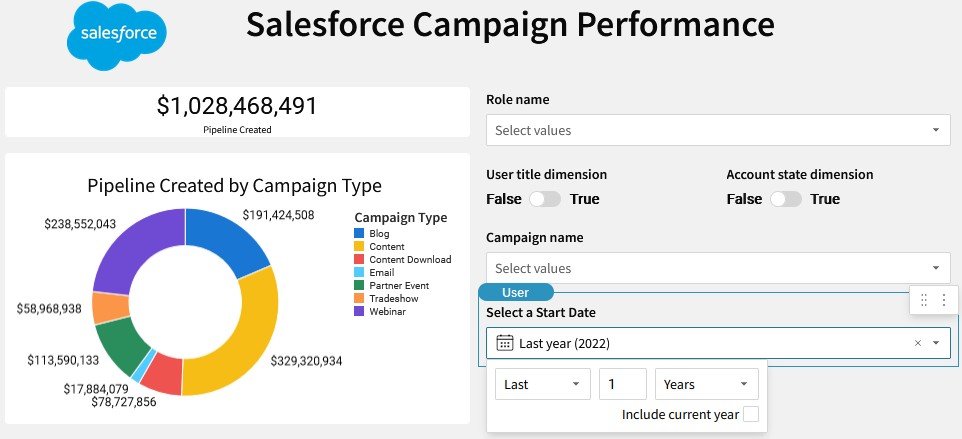}
         \vspace*{-5mm}
         \caption{Dashboard interface for online exploration.}
         \vspace*{-3mm}
        \label{fig:sigmadashboard}
     \end{subfigure}
      \caption{Screenshots for Sigma Computing dashboards.}
      \vspace*{-2mm}
\end{figure}

Dashboards are at the heart of modern BI tools (e.g., PowerBI~\cite{powerbi}, Looker~\cite{looker}, Sigma Computing~\cite{gale2020sigma}) and provide a comprehensive view of a business within a single interface. 
Modern organizations store data across dozens or hundreds of tables in data warehouses, so dashboard creation consists of two stages.
Offline, data engineers pre-define join relationships between relevant tables so that they can be queried like a denormalized ``wide table'', and create dashboard visualizations. 
Online, domain users interact with the dashboards and make business decisions. 
For example, let's consider a hypothetical scenario based on Sigma Computing:

\begin{example}[Sigma Computing Dashboard]
Anna, a sales manager, is responsible for driving revenue growth.  To gain a better understanding of the current state of the business, she asks Shannon, a data engineer, to build a dashboard to display sales pipelines for potential revenue opportunities. Shannon collects relevant tables from Salesforce (e.g., Opportunities, Campaigns, Users) for deals from various sales representatives, and creates the {\bf join graph} $R_{\Join}$ shown in \Cref{fig:sigmajoin}.

Shannon designs the dashboard in \Cref{fig:sigmadashboard}, where each chart is generated from an initial {\bf dashboard query}, and interactions change parts of those queries to update the charts. 
The total "Pipeline Created" is computed by $Q_1{=}\gamma_{SUM(amount)} R_{\Join}$ while the pie chart displaying  "Pipeline Created by Campaign Type" is computed by $Q_2{=}\gamma_{SUM(amount),Campaign\_Type} R_{\Join}$. 
Shannon adds interactions (e.g., dropdown boxes and switches) to filter or change the grouping attributes. 
For example, selecting  ``Sales Associate'' in the ``Role Name'' dropdown triggers an {\bf interaction query} that filters $R_{\Join}$ by the role before aggregating the data for $Q_1, Q_2$,  while toggling the ``User Title'' switch adds the attribute to the group by. 
However, every interaction translates to queries over the join graph that take many seconds to complete, causing Anna to stop using the dashboard.
\end{example}

The pattern in the above example is common.    Data engineers build a dashboard over a complex acyclic join graph.   When a domain user loads the dashboard, it first executes the {\bf dashboard queries} to load the initial view, and then executes many {\bf interaction queries} in response to user manipulations.  Each of the \intqs is similar to the \dashqs, but may add/change a filter, grouping attribute, or add/remove a relation.   The key challenge is that users expect fast response times~\cite{liu2014effects}, yet joins are notoriously expensive to execute~\cite{leis2018query,neumann2018adaptive,albutiu2012massively}. \revise{Traditional techniques, like  cubing~\cite{gray1997data} and indexing~\cite{moritz2019falcon}, are designed for a single table, but poorly handle joins because the denormalized table size can be exponential to the number of relations: $O(n\times f^r)$, where $f$ is the fanout along join graph edges with $r$ relations, each of size $n$.}

Recent factorized query execution techniques~\cite{olteanu2015size,aberger2017emptyheaded} speed up queries over large joins by pushing down aggregation through the joins, in the spirit of projection pushdown. This reduces the space overhead (for acyclic joins) to a linear scale: $O(r n)$, making it promising for developing interactive dashboards. However, naive factorized query execution of  \intqs online still results in high latency. The process requires scanning, joining, and aggregating all the relations as part of the factorized query execution, which prevents achieving interactive speeds.
Recent work~\cite{schleich2019layered} has proposed optimizing a pre-determined batch of factorized queries offline. However, \intqs are only determined by the combination of user interactions online. Batching all possible \intqs offline would lead to a combinatorially large overhead.

In this paper, we present \sys,
a dashboard accelerator for \intqs over large joins.
Offline, the engineering team connects \sys to a DBMS, specifies the join graph (tables and join conditions), and defines Selection-Projection-Join-Aggregation (SPJA) queries (potentially from a BI dashboarding tool) as \dashqs to create visualizations. \sys precomputes compact data structures and stores them as tables in the DBMS; this incurs a constant factor runtime cost relative to running the \dashqs.
Online, \sys supports a wide range of \intqs that modify select/group clauses, update or remove tables, or join new tables to the \dashqs, all at interactive speeds. 

To support efficient aggregation queries over joins, we observe that \intqs differ from the \dashq by keeping the query structure but modifying a subset of the SPJA operators.  To this end, we introduce the novel Calibrated Junction Hypertree (\cjt) data structure to support work sharing and ensure that the work needed to compute an \intq is proportional to {\it how different it is from its corresponding} \dashq (or \intq), rather than its overall complexity.
Our design builds on the observation by Abo et al.~\cite{abo2016faq} that factorized query execution can be modeled as message passing in Probabilistic Graphical Models (PGM)~\cite{koller2009probabilistic}, as described in the following example: 

\begin{figure}
  \centering
     \begin{subfigure}[b]{0.27\textwidth}
         \centering
         \includegraphics[width=\textwidth]{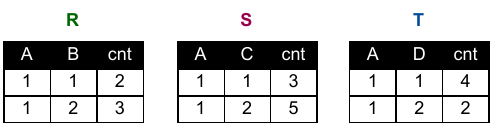}
         \vspace*{-5mm}
         \caption{Relations.}
         \label{fig:relations}
     \end{subfigure}
    \begin{subfigure}[b]{0.2\textwidth}
         \centering
         \includegraphics[width=\textwidth]{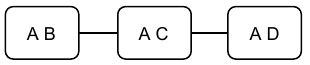}
         \vspace*{-5mm}
         \caption{Join graph.}
         \label{fig:junctionHypertree}
     \end{subfigure}
     \begin{subfigure}[c]{0.4\textwidth}
         \centering
         \includegraphics[width=\textwidth]{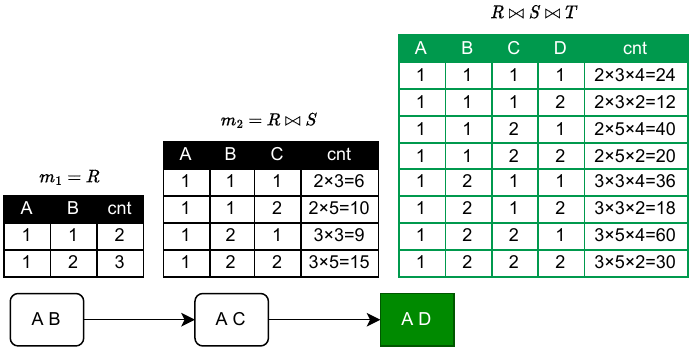}
         \vspace*{-5mm}
         \caption{Naive query execution. The full join result is in green.}
         \label{fig:join}
     \end{subfigure}
     \begin{subfigure}[b]{0.45\textwidth}
         \centering
         \includegraphics[width=0.8\textwidth]{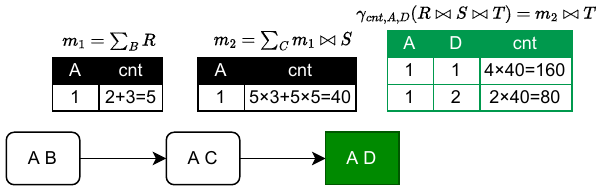}
         \vspace*{-3mm}
         \caption{Upward message passing. The final result is in green.}
         \label{fig:messagepass}
     \end{subfigure}
     \vspace*{-3mm}
     \begin{subfigure}[b]{0.45\textwidth}
         \centering
         \includegraphics[width=0.8\textwidth]{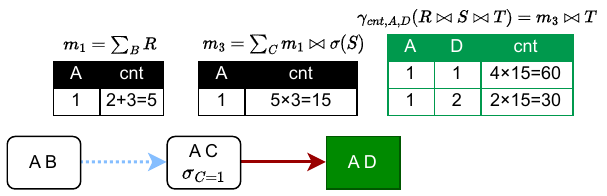}
         \caption{$m_1$ (dotted blue) could be re-used for \intq with additional selection $\sigma_{C{=}1}$ and only $m_3$ (red) need to be computed.}
         \label{fig:messagereuse}
     \end{subfigure}
  \caption{Example database with three relations, its join graph (also \jt), naive query execution for the total count, and factorized query execution by upward message passing.}
  \label{fig:buildDtree}
  \vspace*{-3mm}
\end{figure}

\begin{example}
\label{messagepassingexp}
  \Cref{fig:buildDtree}(a,b) list example relations (duplicates are tracked with a \texttt{cnt} ``annotation'') and the join graph, respectively.  Consider a dashboard query that computes the total count over the full join result: $Q = \gamma_{cnt}(R\Join S \Join T)$.
  \Cref{fig:join} naively executes the query, which computes the full join in order $(R \Join S) \Join T$ before summing the counts, and requires exponential space.
  
  In contrast, factorized query execution distributes the summation through joins, so that each node first sums out (marginalizes) attributes irrelevant downstream, and then emits a smaller message.  Any sequence of messages from leaves to root in the join graph results in the correct result. \Cref{fig:messagepass} chooses $T$ as the root, then passes messages along $R {\to} S {\to} T$. \texttt{AB} marginalizes out \texttt{B} and \texttt{AC} marginalizes out \texttt{C}. Therefore, we only sum 2 tuples to compute the final query result.
\end{example}

The benefit of viewing query execution as message passing is that work-sharing opportunities become self-evident.
Consider the \intq in \Cref{fig:messagereuse}, which adds a predicate $C{=}1$ over $S[AC]$.   Factorized query execution would re-pass messages along $R{\to} S{\to} T$, but misses the opportunity to reuse $m_1$.   
A partial solution is to cache the messages when executing \dashq. However, message contents are sensitive to the message-passing order: if we executed the \dashq along $T{\to} S{\to} R$, then the message between $R$ and $S$ would differ from $m_1$.  Thus, \intqs would be forced to use the same (possibly suboptimal) message passing order, or sacrifice message reuse.  
An alternative is to enumerate all orders and store their messages, but has compute and storage costs quadratic in the number of relations. 

Our algorithmic contribution is to show that after executing the \dashq, sending messages in reverse from root to leaves (e.g., $R {\leftarrow} S {\leftarrow} T$) is sufficient to support any execution order over the join graph.   
We, therefore, use \cjt to proactively materialize \dashq messages for all possible orders.
Such a process is called "calibration" and is first brought up in PGM~\cite{shafer1990probability}, which is used to similarly share computation between queries on posterior distributions.  We are the first to apply it to SPJA queries.

\cjts allow work sharing between \dashqs and \intqs, and we further extend work sharing between \intqs.   The key idea is that, given \dashq $q_0$, after a user performs an interaction (e.g., filter by Role as $q_1$), they may {\it add} another interaction (e.g., further add a filter by Campaign as $q_2$).   In this case, the difference between the successive \intqs ($q_1$ and $q_2$) is smaller than between $q_2$ and $q_0$.   
Since $q_1$ is not known offline, we calibrate it online during ``think-time'' between interactions~\cite{eichmann2020idebench}.  Such calibration doesn't need to be fully completed, and the user's next interaction preempts calibration. The next query can benefit from any messages that are newly materialized.

We implement \sys as a Python dashboard accelerator library. \sys acts as a middleware that transparently rewrites queries from dashboards to benefit from factorized  execution and work sharing from \cjts.   The generated queries are simple and easily ported to different DBMSes or data frame systems~\cite{reback2020pandas,modin}.
Furthermore,  \sys accelerates advanced dashboard features like interactive {\it Data Augmentation}~\cite{chepurko2020arda} for analytics or machine learning.

\smallskip\noindent
To summarize, our contributions are as follows:

\begin{myitemize}
  \item We design the novel \cjt data structure, which improves the efficiency of \intqs over join by reusing messages and applying calibration. The cost of materializing the data structure is within a constant factor of the \dashq execution, but accelerates \intqs by multiple orders of magnitude.

  \item We build \sys, which manages \cjt to transparently accelerate dashboards. It builds \cjts based on initial \dashqs, and uses think time to further calibrate \intqs.   Its simple rewrite-based design is easily portable to any SQL-based DBMS.
  
  \item We evaluate the effectiveness of \sys on both local and cloud DBMSes using real-world dashboards and TPC benchmarks. Our results show that \sys accelerates most \intqs by $>100\times$, and speeds up ML augmentation by   $10{\times}$.
\end{myitemize}

\vspace*{-2mm}

\vspace*{-2mm}
\section{Background}
\label{s:background}
This section provides a brief overview of annotated relations, early marginalization and variable elimination to accelerate join-aggregation queries, and the junction hypertree for join representation.  

\stitle{Data Model.}  
Let uppercase symbol $A$ be an attribute, $dom(A)$ is its domain, and lowercase symbol $a\in dom(A)$ be a valid attribute value. \revise{For the purpose of analytical simplicity, we assume categorical attributes with a fixed domain size.\footnote{However, the system, \sys, doesn't rely on the fixed attribute domain sizes and can trivially support numerical attributes. \sys simply issues SPJA queries to DBMSes (\Cref{s:apps}), which can be executed over relations with numerical attributes. }}.
Given relation R, its schema $S_R$ is a set of attributes, and its domain $dom(R) = \times_{A \in S} dom(A)$ is the Cartesian product of its attribute domains. An attribute is incident of R if $A \in S_R$. Given tuple t, let $t[A]$ be its value of attribute A. 

\stitle{Annotated Relations.} 
Since relational algebra (first-order logic) does not support aggregation, it has been extended with the use of 
commutative structures to support aggregation.  The main idea is that tuples are annotated with values 
from a semi-ring, and when relational operators (e.g., join, project, group-by) concatenate or combine
tuples, they also multiply or add their annotations, such that the final annotations correspond to the aggregation results.

A commutative semi-ring $(D, +, \times, 0, 1)$ consists of a set $D$ and binary operators + and $\times$ that are commutative and closed over $D$, along with the zero $0$ and unit $1$ elements.
\revise{We focus on  semi-rings with elements in $D$ of constant size for efficiency.
The semi-ring structure accommodates nearly all standard aggregates such as count, sum, min, max, etc~\cite{abo2016faq}. More complex aggregates and even ML model can be constructed from semi-ring: variance can be derived from count, and sum ($var(A) = sum(A^2)/count(A) - sum(A)^2/count(A)^2$), and linear regression can be trained based on the sum of pairwise products among features and target variable.
There are two commonly used classes of aggregates
not supported: percentile-based (e.g., median), and distinct-based (e.g., distinct count) aggregates. They require the tracking of a distribution or a unique value set, and cannot be represented by a  set $D$ of constant-sized elements. These can be approximated for future work~\cite{cormode2011sketch}.
}.
For simplicity, the text will be based on \cnt queries and the natural numbers semi-ring $(\mathbb{N},+,\times,0,1)$, which operates as in grade school math.
Each relation $R$ annotates each of its tuples $t\in dom(R)$ with a natural number, and $R(t)$ refers to this annotation for tuple $t$~\cite{green2007provenance,joglekar2015aggregations,nikolic2018incremental}.
We will use the terms {\it relation} and {\it annotated relation} interchangeably.

\stitle{Semi-ring Aggregation Query.}
Aggregation queries are defined over annotated relations, and the relational operators are extended to add or multiple tuple annotations together, so that the output tuples' annotations are the desired aggregated values\footnote{Note that this means different aggregation functions are defined over different semi-ring structures, and our examples will focus on \cnt queries.}.

Consider an example query $\gamma_{\textbf{A},\cnt}(R_1 \Join R_2 ... \Join R_n)$
that joins $n$ relations, groups by a set of  attributes \textbf{A}, and computes the \cnt.
The operators that combine annotations are joins and groupbys 
and they compute the output tuple annotations as follows:
\begin{align}
  (R\Join T)(t) =& \HS R(\pi_{S_R} (t)) \times T(\pi_{S_T} (t)) \\
  (\sum_A R)(t) = & \sum \{R(t_1) | \HS t_ 1 \in S_R , t = \pi_{S_R \backslash  \{A\}} (t_1 )\} 
\end{align}

\noindent (1) states that given a join output tuple $t$, its annotation is the product of counts from the contribution pair of input tuples.
(2) defines the count for output tuple $t$, and $\sum_{A}R$ denotes that we {\it marginalize} over $A$ and remove it from the output schema.  This corresponds to summing the counts for all input tuples in the same group as $t$. \revise{In this paper, we assume natural joins with identical names for join keys for clarity\footnote{For the system, different names can be used as long as join conditions are specified. Our approach can be easily adapted to theta joins and outer joins, by multiplying the annotations of matching tuples (non-existing tuples are annotated by zero)}.}.
To summarize, join and groupby correspond to $\times$ and $+$, respectively. Let the schema of $R_1 \Join R_2 ... \Join R_n$ be \textbf{S}. $q$ can be rewritten as $\sum_{A \in  \textbf{S}-\textbf{A}}(R_1 \Join R_2 ... \Join R_n)$.

\stitle{Early Marginalization.}
In simple algebra (as well as semi-rings), multiply distributes over addition, and can allow us to push marginalization 
through joins, in the spirit of projection push down~\cite{gupta1995aggregate}.  

Consider \Cref{fig:buildDtree}, which computes $\gamma_{A;\cnt}\left(  R\Join S\Join T\right)$.
We can rewrite it as marginalizing $B$, $C$, and $D$ from the full join result 
$\sum_B \sum_C \sum_D R[A,B] \Join S[A,C] \Join T[A,D].$
Although the naive cost is $O(n^3)$ where $n$ is the relation size, we can push down marginalizations: $\sum_D (\blue{\sum_C (}\red{(\sum_B R[A,B])} \blue{\Join S[A,C])} \Join T[A,D])$ where the largest intermediate result, and thus the join cost, is $O(n)$.

\stitle{Join Ordering and Variable Elimination.}
Early marginalization is applied to a given join order. Thus we may also reorder the joins to cluster relations that involve a given attribute, so that it can be safely marginalized.
Consider the query $\sum_{A} R[A,B] \Join S[B,D] \Join T[A,C]$.
We can reorder the joins so that $A$ can be marginalized out earlier: 
$S[B,D] \Join \sum_{A} (R[A,B] \Join T[A,C]).$
The above procedure, where for each marginalized attribute $A$, we first cluster and join relations incident to $A$, and then marginalize $A$, is called variable elimination~\cite{cozman2000generalizing}.
Variable Elimination is widely used for  PGM inference~\cite{koller2009probabilistic} and factorized execution~\cite{abo2016faq}.  
\revise{Variable Elimination reduces the optimization to: (a) identifying the order in which attribute(s) are marginalized out (by clustering and joining the incident relations), and (b) determining the best arrangement of relations within the cluster.
(b) is the traditional join ordering problem~\cite{steinbrunn1997heuristic}. DBMSes that employ binary join use information such as relation cardinality to optimize the ordering.
Prior works~\cite{abo2016faq,schleich2019layered} also apply worst-case-optimal-join (WCOJ~\cite{ngo2018worst}) to simultaneously join these tables for asymptotic improvement.
(a) is called the {\it variable elimination order} and
its complexity is dominated by the intermediate join result size of the clustered relations (using WCOJ). 
It is well known that finding the optimal order (with the minimum intermediate size) is NP-hard~\cite{fischl2018general}. 
However, common DBMS queries are over the acyclic join,  whose optimal order could be found efficiently by GYO-elimination procedure~\cite{yu1979algorithm,abo2016faq}. The central idea is
to repeatedly: (1) eliminate attributes that are present in only one relation, and (2) join relations $R$ with $S$ if the schema of $R$ is a subset of $S$'s schema.}.

\begin{example}
\revise{
Consider $\sum _{ABCD}R[A,B] \Join S[B,C] \Join T[C,D]$, which is acyclic and we apply GYO-elimination: we first eliminate $A, D$ because they each only appear in one relation $R,T$ respectively (the order could also be $D, A$). This results in the intermediates $M_1[B], M_2[C]$, whose schemas are subsets of $S$. So we further join $M_1$, $M_2$, with $S$ and eliminate $B, C$. The final variable elimination order is $ADBC$:
\begin{align*}
&\textstyle\sum _{DBC} S[B,C] \Join T[C,D] \Join \textcolor{red}{\textstyle\sum _{A} (R[A,B])} &\\
=& \textstyle\sum _{BC} S[B,C] \Join M_1[B] \Join \textcolor{red}{\textstyle\sum _{D} (T[C,D])}  &\\
=& \textstyle\sum _{C}  M_2[C] \Join\textcolor{red}{\textstyle\sum _{B} (S[B,C] \Join M_1[B] )} 
=\textcolor{red}{\textstyle\sum _{C} (M_3[C] \Join M_2[C])} 
\end{align*}
Each elimination step is highlighted in \textcolor{red}{red}.  }.
\end{example}

\stitle{Junction Hypertree.} 
The Junction Hypertree\footnote{\jt is also called Hypertree Decomposition~\cite{abo2016faq,joglekar2016ajar}, Join Tree, Join Forest~\cite{idris2017dynamic,schleich2019layered} in databases and Clique Hypertree in PGM~\cite{koller2009probabilistic}.}
(\jt) is a representation of a join query that is amenable to complexity analysis~\cite{abo2016faq,joglekar2016ajar} and semi-ring aggregation query optimization~\cite{aberger2017emptyheaded}. 
Given a join graph $R_1\Join \ldots \Join R_n$, \jt is a pair $(E, V)$, where each vertex $v\in V$ is a subset of attributes in the join graph, and the undirected edges form a tree that spans the vertices.  
The join graph may be explicitly defined by a query, or induced by the foreign key relationships in a database schema.
Following prior work~\cite{abo2016faq}, a \jt vertex is also called a {\it bag}.
A \jt must satisfy three properties:
\begin{itemize}
  \item \stitle{Vertex Coverage:} The union of all bags in the tree must be equal to the set of attributes in the join graph.  
  \item \stitle{Edge Coverage:} For every relation $R$ in the join graph, there exists at least one bag that is a superset of $R$'s attributes.
  \item \stitle{Running intersection:} For any attribute in the join graph, the bags containing the attribute must form a connected subtree.  
\end{itemize}

The last property is important because \jts are related to variable elimination and are used for query execution.  Given an elimination ordering, let each join cluster be a bag in the \jt, and adjacent clusters be connected by an edge.  In this context, executing the variable elimination order corresponds to traversing the tree (path); when execution moves beyond an attribute's connected subtree, then it can be safely marginalized out.  
Note that since the \jt is undirected, it can induce many variable elimination orders (execution plans) from a given \jt, all with the same runtime complexity.

Finally, there are many valid \jt for a given join graph, and the complexity (query execution cost) of a \jt is dominated by the largest bag (the join size of the relations covered by the bag).  Although finding the optimal \jt for an arbitrary join graph is NP-hard~\cite{fischl2018general}, we can trivially create the optimal \jt for an acyclic join graph by creating one bag for each relation (e.g., the \jt is simply the join graph) and the size of each bag is bounded by its corresponding relation size. We refer readers to FAQ~\cite{abo2016faq} for a complete description.

\stitle{Message Passing for Query Execution.}
Message Passing was first introduced by Judea Pearl in 1982~\cite{pearl1982reverend} (known as belief propagation) in order to efficiently perform inference (compute marginal probability) over probabilistic graphical models. 
In database terms,  each probability table corresponds to a relation, the probabilistic graphical model corresponds to the full join graph in a database (as expressed by a \jt), the joint probability over the  model  corresponds to the full join result, and marginal probabilities correspond to grouping over different sets of attributes.
To further support semi-ring aggregation, Abo et al.~\cite{abo2016faq} established the equivalence between factorized query execution, and (upward) message passing. The full algorithm can be found in \Cref{messagepassingalg};
below, we illustrate how message passing over \jt is used for query execution.

The procedure first determines a traversal order over the \jt---since the \jt is undirected, we can arbitrarily choose any bag as the root and create directed edges that point towards the root---and then traverses from leaves to root.  We first compute the initial contents of each bag by joining the necessary relations based on the bag's attributes.  When we traverse an outgoing edge from a bag $l$ to its parent $p$, we marginalize out all attributes that are not in their intersection---the result is the {\it Message} between $l$ and $p$.  The parent bag then joins the message with its contents.  Each bag waits until it has received messages from all incoming edges before it emits along its outgoing edge. Once the root has received all incoming messages, its updated contents correspond to the query result.

\begin{example}[Message Passing]
Consider the relations in \Cref{fig:relations}, and the \jt in \Cref{fig:junctionHypertree} where each bag is a base relation.
  We wish to execute $\sum_{ABCD} R(A,B) \Join S(A,C) \Join T(A,D)$ by 
  traversing along the path $R{\to} S{\to }T$ (\Cref{fig:messagepass}).
  We first marginalize out $B$ from $AB$, so the message to $AC$ is a single row with count $5$.   
  The bag $AC$ joins the row with its contents, and thus multiplies each of its counts by $5$.
  It then marginalizes out $C$, so its message to $AD$ is a single row with count $(3+5)\times 5$.
  Finally, bag $AD$ absorbs the message (\Cref{fig:messagepass}) and marginalizes out $A$ and $D$ to compute the final result.
\end{example}

\section{Calibrated Junction Hypertree}
\label{cjt_detail}
While message passing over \jt exploits early marginalization to accelerate query execution, it has traditionally been limited to  
single-query execution. 
This section introduces Calibrated Junction Tree (\cjt) to enable work-sharing for interactive dashboards on large joins.  
The idea is to materialize messages over the \jt for \dashq, and reuse a subset of its messages for \intqs.  
This section will focus on the basis for the \cjt data structure and how it is used to execute \intqs. The next section will describe how \sys applies \cjt to build an interactive dashboard.  

Our novelty is (1) to use \jts as a concrete data structure to support message reuse, and (2) to borrow {\it calibration}~\cite{shafer1990probability} from PGM to materialize messages for any message passing order.   \revise{Although \cjt is widely used across engineering~\cite{zhu2015junction,ramirez2009fault}, ML~\cite{braun2016lifted,deng2014large}, and medicine~\cite{pineda2015novel,lauritzen2003graphical}, it was used only for probabilistic inference (sum over probability); we are the first to extend it to general SPJA queries with semi-ring aggregations in DBMS for work sharing.}.

\subsection{Motivating Example}
\label{sec:mot}
We illustrate the work sharing between 
a \dashq $Q_1 = \sum_{ABCD} R(A,B) \Join S(A,C) \Join T(A,D)$, and an \intq $Q_2 = \sum_{ABCD} R(A,B) \Join \blue{\sigma_{C=1}}(S(A,C)) \Join T(A,D)$ with additional predicate \texttt{C=1}, to motivate \cjt.

\begin{figure}
  \centering
  \begin{subfigure}[t]{0.23\textwidth}
         \centering
         \includegraphics[width=0.9\textwidth]{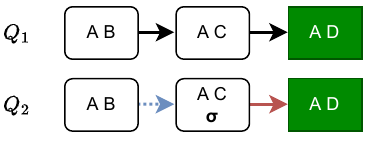}
         \vspace*{-1mm}
          \caption{Message passing to root AD.}
          \vspace*{-1mm}\label{fig:sharebetweenqueries}
     \end{subfigure}
     \hfill
    \begin{subfigure}[t]{0.23\textwidth}
         \centering
         \includegraphics[width=0.9\textwidth]{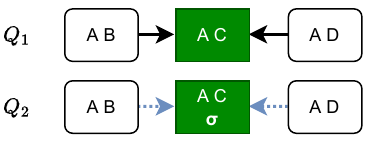}
         \vspace*{-1mm}
         \caption{Moving root increases reuse.}
          \vspace*{-1mm}
          \label{fig:sharebetweenqueries2}
     \end{subfigure}
     \hfill
     \vspace*{-3mm}
      \caption{Work sharing between queries $Q_1$ (total count query) and $Q_2$ (additional selection to S). Dotted \blue{blue edges} are reusable messages and solid \red{red edges} are non-reusable. }
      \vspace*{-4mm}
\end{figure}

\begin{example}
 Consider the \jts in \Cref{fig:sharebetweenqueries} which assign $AD$ as the root for $Q_1, Q_2$ and traverse along the path $R\to S\to T$. 
  Although the message $R\to S$ will be identical (\blue{blue} edges), the additional filter over $S$ means that its outgoing message (and all subsequent messages) will differ from $Q_1$'s and cannot be reused (\red{red} edges).
In contrast, \Cref{fig:sharebetweenqueries2} uses $S$ as the root, so both messages can be reused and the 
$S$ bag simply applies the filter after joining its incoming messages.
\end{example}

This example shows that message reuse depends on how the root bag is chosen for \dashq ($Q_1$),  
and for different \intqs, we may wish to choose different roots.  
Since we don't know the exact join, grouping, and filter criteria of future \intqs, the naive solution is to (costly) materialize messages for all possible roots.
We next present \cjt, a novel data structure for query execution and message reuse, and address these limitations.

\begin{figure}
  \centering
      \includegraphics [scale=0.65] {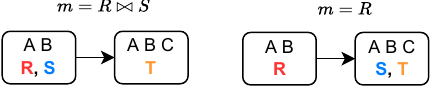}
      \vspace*{-3mm}
      \caption{ The same \jt over relations \textcolor{red}{R(A,B)}, \textcolor{blue}{S(A)}, \textcolor{orange}{T(B,C)} can have different relation mappings ($\mathcal{X}$) and each mapping results in different messages ($m$). For each bag, its attributes are at the top and mapped relations are at the bottom.}
      \vspace*{-3mm}
  \label{fig:diffmsg}
\end{figure}

\subsection{Junction Hypertree as Data Structure}
\label{sec:jtdatastructure}

A naive approach to re-use messages is to execute an aggregation query over \jt, and cache the messages; when a future query traverses an edge in the \jt, it reuses the corresponding message.  Unfortunately, this is 1) inaccurate, because messages generated along an edge are not symmetric and depend on the specific traversal order during message passing, 2) insufficient, because it cannot directly express filter-group-by queries, and 3) leaves performance on the table.    
To do so, we extend \jt as follows:

\stitle{Directed Edges.} To support arbitrary traversal orders, we replace each undirected edge with two directed edges, and use $\mathcal{Y}(i\to j)$ to refer to the cached message for the directed edge $i\to j$.

\stitle{Relation Mapping.} $\mathcal{X}(R)$ maps each base relation $R$ to exactly one bag containing $R$'s schema.  Although different mappings can lead to different messages (\Cref{fig:diffmsg}), acyclic join graphs have a good default mapping where each bag is mapped by a single relation.
Relations mapped to the same bag are joined during message passing.

\begin{figure}
  \centering
  \begin{subfigure}[t]{0.23\textwidth}
         \centering
         \includegraphics[width=0.7\textwidth]{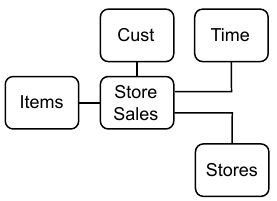}
         \vspace*{-2mm}
          \caption{TPC-DS join graph (also \jt)}
          \label{fig:tpc_ds_schema}
     \end{subfigure}
     \hfill
    \begin{subfigure}[t]{0.23\textwidth}
         \centering
         \includegraphics[width=0.7\textwidth]{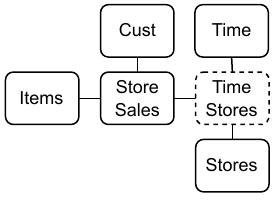}
         \vspace*{-2mm}
         \caption{Add empty bag (Time, Stores).}
          \label{fig:tpc_ds_empty}
     \end{subfigure}
     \hfill
     \vspace*{-3mm}
      \caption{Simplified Join graph (\jt) of TPC-DS. Adding an empty bag can accelerate queries group-by \texttt{Time} and \texttt{Stores}.}
      \vspace*{-3mm}
\end{figure}

\stitle{Empty Bags.}  To avoid large paths during message passing, it's beneficial to add custom {\it empty} bags to create ``short cuts''.  {\it Empty} bags are not mapped from any relations and are simply a mechanism to materialize custom views for work sharing.  They join incoming messages, marginalize using standard rules, and materialize the outgoing messages.   Empty bags are a novel addition in this work:  previous works~\cite{abo2016faq,aberger2017emptyheaded,xirogiannopoulos2019memory} focus on non-redundant \jt without empty bags. This is because they are in the context of single query optimization, where empty bags offer no advantage. 
\begin{example}[Empty Bag]
  Consider the simplified TPC-DS \jt in \Cref{fig:tpc_ds_schema}. Store Sales is a large fact table (2.68M rows at SF=1), while the rest are much smaller.  To accelerate a query that aggregates \texttt{sales} grouped by \texttt{(Store,Time)}, we can create the empty bag \texttt{Time Stores} between \texttt{Store\_Sales}, \texttt{Time} and \texttt{Stores} (\Cref{fig:tpc_ds_empty}). The message from \texttt{Store\_Sales} to the empty bag is sufficient for the query and is $17.3\times$ smaller (154K rows) than the fact table.
\end{example}

Note that leaf empty bag may result in an empty output message; we avoid this special case by mapping the identity relation $\mathcal{I}$\footnote{The schema is the same as the bag and all tuples in its domain are annotated with 1 element in the semiring.} to it, such that $R\Join \mathcal{I} = R$ for any relation $R$. Essentially, the empty bag is ``pass-through'' and doesn't change the join results nor the query result. When the bag is a leaf node, its message is simply $\mathcal{I}$.  We do not materialize the identity relation, as it's evident from the \jt.

\begin{figure}
  \centering
     
    \begin{subfigure}[c]{0.2\textwidth}
         \centering
         \includegraphics[width=\textwidth]{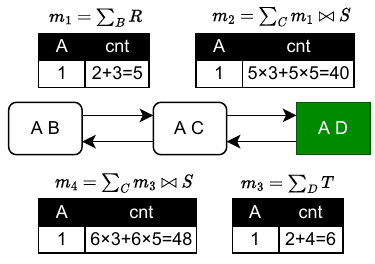}
         \caption{Upward and downward message passing.}
         \label{fig:downward}
     \end{subfigure}
     \hfill
     \begin{subfigure}[c]{0.24\textwidth}
         \centering
         \includegraphics[width=\textwidth]{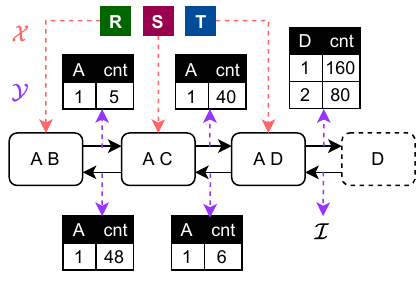}
         \caption{Calibrated Junction Hypertree.}
         \label{fig:datastructure}
     \end{subfigure}
     \hfill
     \vspace*{-3mm}
  \caption{Message Passing and Calibration.  Green rectangle is the root. Dotted one is the empty bag. $\mathcal{I}$ is identity relation. $\mathcal{X}$ maps relations to bags, and $\mathcal{Y}$ maps edges to messages.}
  \vspace*{-3mm}
\end{figure}

\begin{example}[\jt Data Structure]
 \Cref{fig:datastructure} illustrates the \jt data structure for the example in \Cref{fig:buildDtree}. Each relation maps to exactly one bag (orange dotted arrows), and each directed edge between bags (black arrows) stores its corresponding message (purple dashed arrows).  Bag D (dotted rectangle) is an empty bag and materializes the view of "count group by D".  $\mathcal{I}$ is the identity relation.
\end{example}

\begin{table*}
\begin{center}
\setlength{\tabcolsep}{0.4em} 
\begin{tabular}{  m{5em}  m{31em} m{12em} l} 
  \textbf{Annotation} & \textbf{Effect} & \textbf{Applicability} & \textbf{Section}   \\ 
  $\gamma_A$ & Prevent A from being marginalized out for all downstream bags. & Any bag containing A. & \Cref{sec:msgpassing}  \\ 
  $\sum_A$ & Marginalize out $A$.  ``Cancels'' $\gamma_A$ for downstream bags. & Any bag containing A. & \Cref{sss:cjtqexec} \\ 
$\overline{R}$ & Exclude relation R from the bag during message passing. & The bag $\mathcal{X}(R)$.  & \Cref{sec:msgpassing}\\ 
$R^*_{ver.}$. & Update relation R in the bag to the specified version during message passing. & The bag $\mathcal{X}(R)$.  & \Cref{sec:msgpassing}\\ 
$\sigma_{id}$ & Apply selection uniquely identified by $id$ to relations during message passing.  & Any bag covers reference atts.& \Cref{sec:msgpassing}  \\ 
\end{tabular}

\end{center}
\caption{Table of annotations, their effects and applicability.}
\vspace*{-9mm}
\label{table:annotations}
\end{table*}

\begin{figure}
  \centering
         \includegraphics[width=0.35\textwidth]{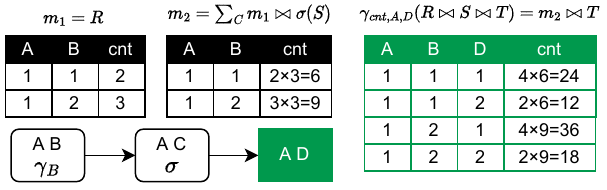}
         \vspace*{-3mm}
         \caption{Filter-group-by query with annotated \jt.}
         \vspace*{-3mm}
         \label{fig:group_by_filter_query}
\end{figure}

\subsection{Message Passing Over Annotated Bags}
\label{sec:msgpassing}

We now describe support for general SPJA queries over \jt.  Although each query \jt has the same structure, we annotate the bags based on the query's SPJA operations. We then modify message passing rules to accommodate the bag annotations.    These annotations will come in handy when determining work-sharing opportunities for a new interactive query given a \dashq.  

Given the database  $\textbf{R}= \{R_1,R_2,...,R_n\}$ and \jt $ = ((E,V), \mathcal{X}, \mathcal{Y})$, we focus on semi-ring SPJA queries of the following form:\\
\indent\texttt{SELECT \textcolor{red}{$\mathcal{G}$}, COUNT(*)} \texttt{FROM \textcolor{blue}{$\mathcal{J}$}}\\
\indent\texttt{WHERE [JOIN COND] AND \textcolor{orange}{$\mathcal{P}$}} 
\texttt{GROUP BY \textcolor{red}{$\mathcal{G}$}}\\
\noindent where $\mathcal{G}$ is the grouping attributes, $\mathcal{J}\subseteq \mathbf{R}$ is the set of relations joined in the \texttt{FROM} clause, and $\mathcal{P}$ is the set of predicates referencing attributes in one bag. 
Query execution is based on message passing (\Cref{s:background}). However, the processing of each bag differs based on annotations.
We propose 4 annotation types, summarized in \Cref{table:annotations}:

\begin{myitemize}

\item \texttt{\textcolor{red}{GROUP BY $\mathcal{G}$.}} For each attribute $A\in\mathcal{G}$, we annotate exactly one bag $u$ that contains this attribute with $\gamma_A$.  Messages emitted by the annotated bag and all downstream bags do not marginalize out $A$.  Since all bags containing $A$ form a connected subtree (running intersection), which bag we annotate does not affect correctness. However, we will later discuss the performance implications of different choices when we use  \cjt to queries.

\item \texttt{\textcolor{blue}{Joined Relations $\mathcal{J}$.}} The query may not join all relations  in the join graph, or the joined relations are updated.
  For each relation R {\it not} included (resp. updated) in the query, we annotate the corresponding bag $u=\mathcal{X}(R)$  with $\overline{R}$ (resp. $R^*_{ver.}$). When computing messages from this bag, R will be excluded from $\mathcal{X}^{-1}(u)$\footnote{Rigorously, $\mathcal{X}$ doesn't have an inverse function. We define $\mathcal{X}^{-1}$ to be a mapping from one bag to a set of base relations such that $\mathcal{X}^{-1}(u) = \{i| \mathcal{X}(i) = u\}$.} (resp. R will be updated in $\mathcal{X}^{-1}(u)$). We allow only the exclusions of relations that don't violate \jt properties. 

\item \texttt{\textcolor{orange}{PREDICATES $\mathcal{P}$.}} Let predicate $\sigma\in\mathcal{P}$ be over attribute $\mathbf{A}$.  
  We choose a bag $u$ such that $\mathbf{A} \subseteq u$, and annotate it with $\sigma_{id}$---the effect is that the predicate filters all messages emitted by $u$.   
  The choice of bag to annotate is important---for a single query, we want to pick a bag far from the root in the spirit of selection push down, whereas to maximize message re-usability, we want to pick the bag near the root.  We discuss this trade-off in \Cref{sss:msgreuse}.

\end{myitemize}

\subsubsection{Message Passing}
\label{sss:msgpassing}

We now modify how message passing, generation, and absorption work to take the annotations into account.

\stitle{Upward Message Passing.} 
Traditional message passing chooses a root bag and traverses edges from leaves to the root.  Since  \jt uses bidirectional edges, we call this "upward message passing".
The message $\mathcal{Y}(b{\to}p)$ from bag $b$ to parent $p$ is defined as follows:
Let $M(b,p) = \{\mathcal{Y}(i{\to}b) | i{\to}b{\in} E\land i{\not\eq} p\}$ be the set of incoming messages (except from $p$).
We join between all  relations (updated to the specified versions) in $M(b)$ and $\mathcal{X}^{-1}(b)$, and marginalize out all attributes not in $p$.
Given annotations, we exclude relations in $\overline{R}$ from the join, apply predicates $\sigma$ (with appropriate push-down), and exclude attributes in $\gamma$:
$\mathcal{Y}(b\to p) = \textstyle\sum_{b-(p\cap b)-\gamma} \sigma(\Join (M(b,p) \cup \mathcal{X}^{-1}(b) - \overline{R}))$. $b$'s message to $p$ is ready iff all its messages from child bags are received.
During message passing, if $b$ contains group-by annotation $\gamma$,  we temporarily annotate all its downstream bags also with $\gamma$.

\stitle{Absorption.} 
Absorption is when the root bag $r$ consumes {\it all} incoming messages.
It is identical to the join and filter during message generation:
$Absorption(r) {=} \sigma(\Join (M(r,\varnothing) \cup \mathcal{X}^{-1}(r) - \overline{R}))$.
To generate the final results, we marginalize away all attributes not in $\mathcal{G}$.

\begin{example}
Consider database and \jt in \Cref{fig:buildDtree}. Suppose we want to query the total count filter by C = 1 and group by B. This requires us to annotate bag AB with $\gamma_B$ and bag AC with $\sigma$ (id is omitted). \Cref{fig:group_by_filter_query} shows the upward message passing over the annotated \jt to root AD, where attribute B is not marginalized out and the predicate $C {=} 1$ is applied to S. After upward message passing, bag AD performs absorption and marginalizes out AD to answer the query. 
\end{example}

\subsubsection{Runtime Complexity}
\label{sss:runtimecomp}

\revise{
The main purpose of this runtime complexity analysis is to serve as a baseline for analyzing the runtime benefits of the \cjt in \Cref{sec:benefit}. Our analysis largely follows FAQ~\cite{abo2016faq}, with extensions to consider the effects of selection (by selectivity) and group-by (by attribute domain size) annotations.
While FAQ also supports group-by, it's by modifying variable elimination to construct a new \jt, and bounding \jt by relation sizes. 
We first use an example to contrast the difference:}.
\vspace*{-2mm}
\begin{example}
\revise{Consider relations $R(A,B), S(A,C)$ with a \jt of $(AB)-(AC)$; each relation and attribute domain size is $O(n)$ and $O(d)$ respectively. For the query $\gamma_{B, C, count(*)}(R\Join S)$, \sys annotates $(AB)$  with $\gamma_B$ and $(AC)$ with $\gamma_C$, leading to a \jt of $(AB\gamma_B)-(AC\gamma_C)$.  If $(AC)$ is the root, the absorption of $(AC)$ is bounded by $O(nd)$ (increased by $O(d)$ for $\gamma_B$).
But in FAQ, this is viewed as variable elimination of only $A$, leading to \jt of a single bag $(ABC)$ with absorption size $O(n^2)$.}.
\end{example}
\vspace*{-1mm}
\revise{Our extension offers two benefits: 
(1) it improves message reuse as compared to  running variable elimination to construct a new \jt, and (2) dashboards typically have few group-by attributes with domain sizes much smaller than relation sizes, which improve the bounds\footnote{If there are many group-by attributes with large domains,  the bound can be tightened by fractional edge cover based on the relation sizes and functional dependency~\cite{abo2016computing}.}. Next, we lay out the setting for our analysis:}.
\revise{\begin{myitemize}
  \item {\bf Measure:} We measure complexity in terms of both the query and database sizes using the standard RAM model of computation.  
  \item {\bf Query:} We focus on a SPJA query over an acyclic\footnote{Or cycles has been pre-joined, following standard hypertree decomposition ~\cite{abo2016faq,joglekar2016ajar}.} natural join  of $n$ relations $R_1, ..., R_n$ with attribute domains of size $O(d)$. The selections and group-bys have been annotated in \jt (below).
  \item {\bf Annotated \jt:} Let the annotated $\jt(E,V,\mathcal{X},\mathcal{Y})$ over $n$ bags be the SPJA query. Each bag maps to exactly one relation, whose schema is the same as bag attributes. To simplify notation, we denote the relation of bag $u$ as $R_u = \mathcal{X}^{-1}(u)$. For annotations:
  \begin{myitemize}
  \item {\bf Selection:} $s(u)$ denotes the combined selectivity of all $\sigma$ on $u$. 
  \item {\bf Group-by:} $g(u, v)$ denotes the number of attributes to group-by from the upstream bags to $u\to v$ (excluding the group-bys in $u$).  Note that the $\sum$ annotations can cancel out the corresponding group-bys and are not counted in $g(u,v)$.
  \item {\bf Update:} We assume $R_u$ references the latest relation.
  \item {\bf Exclusion:} While relation exclusion is common for multi-relation bag (e.g., for graph analytics~\cite{robinson2015graph}), it's rare for single-relation bag, and may lead to disconnected join graph, Cartesian products, and size blowup  (except for leaf bag). For simplicity, we assume no relation exclusion.
  \end{myitemize}
  \item {\bf Query Execution:} Following FAQ~\cite{abo2016faq}, we analyze the WCO LeapFrog Triejoin~\cite{veldhuizen2012leapfrog}. Given a natural join of $n$ relations, $m$ unique attributes, maximum input relation size $N$, and the fractional edge cover bound~\cite{atserias2008size} of join size $\rho$ (based on the join graph and relation sizes), the runtime complexity is $O(m n{\cdot}\rho{\cdot}log N)$.
  For SPJA, we pre-apply selection, join, and use standard hash-based aggregation in $O(\rho)$; join dominates the aggregation cost.
\end{myitemize}}.
\vspace*{-2mm}
\revise{\begin{prop}[Runtime Complexity] 
\label{prop:messagereuse}
Executing  SPJA query naively takes $O(|V| \cdot |\cup_{v\in V} v| \cdot \rho \, log(max_{v\in V} s(v)\cdot|R_v|) + \sum_{v\in V} |R_v|)$, where  $\rho$ is the fractional edge cover bound of join size over selected relations. For message passing over \jt, given root $r \in V$, define $Tra(r, \jt)$ as the set of pairs $(u,v)$ for all directed edges $u{\to}v$ on the path to $r$ in $E$.
The complexity is then 
$O(
\textstyle\sum_{u, v \in Tra(r,\jt) \cup \{(r,\varnothing)\}}\mathcal{M}(u,v)
+|R_u|)$, where $\mathcal{M}(u,v) {=} (|U(u,v)|+1)(|u| + g(u, v)) \mathcal{S}(u,v)\, log\, 
max(max_{x \in U(u,v)}\mathcal{S}(x,u),\\ s(u)|R_u|)$ is the message passing ($u{\to}v$) cost, $U(b,p) = \{i | i{\to}b\in E\land i\not\eq p\}$ is the set of $b's$ neighbour bags  (except $p$) and $\mathcal{S}(u,v) = d^{g(u, v)}s(u)|R_u|$ is the size bound of message from $u$ to $v$.
\end{prop}}.
\vspace*{-3mm}
\revise{\begin{proofk}
The naive SPJA query execution runtime complexity is the sum of leapfrog triejoin  and selection cost (in $O(\sum_{v\in V} |R_v|)$).
Message passing analysis is similar with one difference: messages are aggregated results. This allows us to bound the join output sizes by the domain size of group-by attributes.
For each message $u\to v$, we join $u$ with all its incoming messages and aggregate; the cost is dominated by the join. Without group-by annotation, the join size is bounded by $O(s(u)|R_u|)$\footnote{The effects of selections from upstream bags are not accounted for simplicity; standard cost estimation~\cite{selinger1979access} can approximate the combined selectivity for optimization.}.  
With group-by annotations, the join (and message) size increases by a factor of the group-by attributes' domain size $O(d^{g(u, v)})$
Absorption for $r$ follows the same analysis. The final complexity sums selection, message passing, and absorption.
\end{proofk}}.

\revise{Beside the polynomial difference in query complexity, join sizes are the core difference in data complexity:  naive query execution materializes joins ($O(\rho)$), which can be prohibitive. In contrast, factorized execution restricts the join size to $d^{g(u, v)}s(u)|R_u|$ for each message passing $u\to v$, which in practice is much smaller. }.

\subsubsection{Single-query Optimization.} For a given SPJA query, we can choose different bags to annotate, and different roots for upward message passing. We make these choices based on heuristics that minimize the runtime complexity.  Since relation removal and update annotations $\overline{R},R^*_{ver.}$ can be only placed on $\mathcal{X}(R)$, and the placement of group-by don't affect the message passing, the only factor is the choice of root bag and selection annotations.  We enumerate every possible root bag, greedily push down selections, and choose the root with the smallest complexity; the total time complexity to find the root is polynomial in the number of bags.

\subsubsection{Message Reuse Across Queries} 
\label{sss:msgreuse}
Messages reuse between queries requires that the message along edge $u\to v$ only depends on the annotated sub-tree rooted at $u$.  Thus, a new query can reuse materialized messages in \cjt that have the same subtree (and annotations).  

\begin{prop}[Message Reusability]
\label{prop:messagereuse}
Given a \jt and annotations for two queries, consider the directed edge $u\to v$ present in both queries.  Let $T_u$ be the subtree rooted at $u$. If the annotations for $T_u$ are the same  for both queries, then the message along $u\to v$ will be identical irrespective of the traversal order nor  choice of the root.
\end{prop}

This proposition is well established in PGM~\cite{shafer1990probability}, and follows for message passing over \jt.    The proof sketch is as follows: leaf nodes send messages that only depend on outgoing edges, base relations and annotations, while a given bag's outgoing message only depends on its mapped relations ($\mathcal{X}$), annotations and  incoming messages. None of these depend on the traversal order nor the root.

\Cref{prop:messagereuse} implies that an annotation can ``block'' reuse along all of its downstream messages. For group-by annotation, we greedily push down it to the leaf of the connected subtree closest to the root to maximize reusability.
However, pushing selections down trades-offs potentially smaller message sizes for limited reusability; we discuss this interesting optimization problem in \Cref{s:selectionopt}.

\subsection{Calibration}
\label{sec:calibration}
We saw above that message reuse depends on choosing a good root for message passing, however upward message passing only materializes messages for a single root.  \textit{Calibration} materializes messages for all roots, letting future queries pick arbitrary roots.

\subsubsection{Calibration} 
Given an edge $u\to v$, $u$ and $v$ are calibrated iff their marginal absorption results are the same in both directions: 
$\sum_{u - (v \cap u)}  Absorption(u) = \sum_{v - (v \cap u)} Absorption(v)$.
The \jt is calibrated if all pairs of adjacent bags are calibrated. We call this a {\it Calibrated Junction Hypertree} (\cjt), which is achieved by Downward Message Passing discussed next.

\stitle{Downward Message Passing}. Upward message passing computes messages along half of the edges from leaves to root. Downward Message Passing simply reverses the edges and passes messages from the root (now the leaf) to the leaves (now all roots).  Now, all directed edges store materialized messages.

\begin{example}
Consider the example in \Cref{fig:downward}. During upward message passing, root $AD$ receives the message from leaf $AB$. After that, we send messages back from $AD$ to $AB$. We can verify that the \jt is calibrated by checking the equality between the absorptions.
\end{example}

Calibration means all bags are ready for absorption.
This immediately accelerates the class of queries that furthers adds one grouping or filtering over attribute $A$. We simply pick a bag containing $A$ and apply the filter/group-by to its absorption result.

\subsubsection{Query Execution Over a \cjt}
\label{sss:cjtqexec}

How do we execute a new interactive query $Q$ over the \cjt of a \dashq $Q_p$?   
Since they share the same \jt structure, they only differ in their annotations.  The main idea is that query execution is limited to the subtree where the annotated bags differ between the two queries, while we can reuse messages for all other bags in the \cjt.

\begin{figure}
  \centering

         \includegraphics[width=0.45\textwidth]{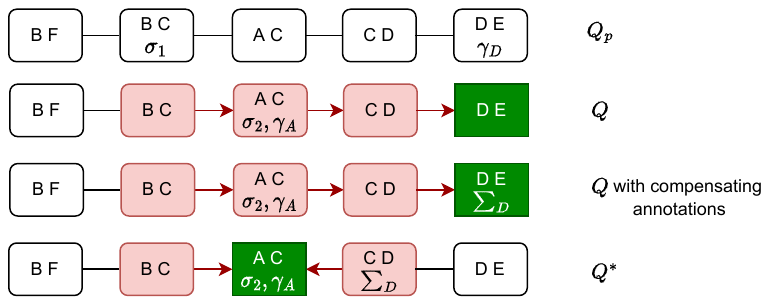}
         \vspace*{-3mm}
         \caption{Given \cjt of $Q_p$, the Steiner tree to execute $Q$ is highlighted (green is root and red is non-root nodes). Compensating annotation $\sum_D$ is introduced to compensate $\gamma_D$, and can be moved for better plan.
         $Q^*$ is the optimal query with minimum Steiner tree size and runtime complexity. }
         \vspace*{-2mm}
         \label{fig:steiner_tree}

\end{figure}

Let \AnnoP and \Anno be the set of annotations for $Q_p$ and $Q$, respectively; note that the annotations in \AnnoP are bound to specific bags in the \cjt, while the annotations in \Anno are not yet bound.  Further, let \BagDiff be the subset of bags whose annotations differ between the two queries.  The minimal {\bf Steiner tree} $T$ is the subtree in the \cjt that connects all bags in \BagDiff using the least number of bags.
Owing to the simplicity of tree structure, the minimal Steiner tree for a specified set of \BagDiff can be identified efficiently.
From \Cref{prop:messagereuse}, edges that cross into $T$ have the same messages as in the \cjt and can be reused.  Thus, we can only perform upward message passing inside of $T$.    Let us first start with an illustrative example:

\begin{example}[Steiner Tree]
  In \Cref{fig:steiner_tree}, the \dashq $Q_p$ groups by \texttt{D}  and filters by $B = 1$, and so its annotations are \AnnoP $= \{\sigma_1, \gamma_D\}$. 
  Suppose query $Q$ (row 2) instead groups by A and filters by $C = 1$ (\Anno $= \{\sigma_2, \gamma_A\}$), and we place its annotations $\sigma_2$ and $\gamma_A$ on \texttt{AC}. The two queries differ in bags \BagDiff=$\{BC, AC, DE\}$, and we have colored their Steiner tree.  Therefore, we can reuse the message $BF\to BC$, but otherwise re-run the upward message passing along the Steiner tree.

\end{example}

Although the example allows us to reuse one message, it's sub-optimal because \BagDiff can be further compressed, and the root is poorly chosen.  Instead, we use a greedy procedure: we arbitrarily place the annotations on valid bags to create an initial Steiner tree, and then greedily shrink it.  Given the minimal Steiner tree over shrinked the \BagDiff, we find the optimal root following \Cref{sec:msgpassing}. 

\stitle{Initialization.}
For annotations only in \Anno, they are added to $Q$'s \jt based on the single-query optimization rules in \Cref{sec:msgpassing}.
For annotations only in \AnnoP, we need to compensate for their effects.  For $\sigma_p$ and $\overline{R}$, we remove the annotation, while for $\gamma_D$, we introduce the {\it compensating annotation} $\sum_D$, which marginalizes out $D$, and place it on the same bag.  A unique property of $\sum_D$ is that we can freely place it on any bag that contains $D$.    For all of the above annotations, we add their bags to \BagDiff.  This defines the initial Steiner tree:
\vspace*{-2mm}
\begin{example}
  The third row in \Cref{fig:steiner_tree} adds the compensating annotation $\sum_D$ to \texttt{DE}.  Its execution is as follows: \texttt{BC} doesn't apply $B=1$, \texttt{AC}  applies $C=1$, groups by $A$, and \texttt{DE} maginalizes out $D,E$.  
\end{example}
\vspace*{-2mm}
\stitle{Shrinking.}
Given the leaves of the Steiner tree, we try to move the differed annotations of $Q$ toward the interior of the tree.  Recall that $\sigma$, $\gamma$, and $\sum$ can be placed on any bag containing the annotation's attribute.  We greedily choose the bag with the largest underlying relation and move its annotations first to reduce the Steiner tree.  
\begin{example}
  $Q^*$ in \Cref{fig:steiner_tree} shows the optimal execution plan over the minimal Steiner tree for $Q$.   It has moved $\sum_D$ to \texttt{CD}, and made \texttt{AC}  the root.   \texttt{CD} will marginalize out \texttt{D}, and \texttt{AC} performs the filter and group-by. In this way, we also reuse the message $\texttt{DE} \to \texttt{CD}$. 
\end{example}
\vspace*{-5mm}
\revise{\subsubsection{Runtime Benefit.}
\label{sec:benefit}
After optimization, the query execution runtime complexity with \cjt is primarily 
dictated by the difference between the current query and the \cjt, as measured by the size of the Steiner tree, and is never worse than naive execution without the \cjt.
The proof sketch is as follows: Consider naive query execution over $\jt(E,V,\mathcal{X},\mathcal{Y})$ with root $r$.
With \cjt, we can share the message outside the Steiner tree $\steiner(E'\subseteq E,V'\subseteq V)$. We analyze the simple case of message passing over \cjt with the original annotation placements {\it without shrinking} and the same root choice $r$, which is sufficient to show the runtime benefit; better root choices and shrinking can further improve the time complexity. Let $r'=r$ if $r\in V'$; otherwise, $r'$ is  the bag  $\in V'$ closest to $r$. The runtime complexity for query execution with \cjt with root $r$ becomes:
$$O(
\textstyle\sum_{u, v \in (Tra(r,\jt)  \cup \{(r,\varnothing)\}) \textcolor{red}{ \setminus  (Tra(r',\jt)  \setminus  Tra(r',\steiner))}}\mathcal{M}(u,v)+|R_u|)$$
The core benefit of \cjt is to enable the sharing of messages highlighted in \textcolor{red}{red}.
These messages have all upstream bags outside of \steiner, whose annotations are thus the same. They can therefore be shared according to \Cref{prop:messagereuse}.
As we can see, similar queries require fewer annotations, and thus a smaller Steiner tree that requires the same or fewer messages.
The benefits are particularly pronounced for imbalanced relation sizes, such as a snowflake schema, where changes are in the small dimension tables.   Messages from the fact table can be shared to avoid joining and aggregating the fact table.  }

\stitle{Applying \cjt to Dashboard and Challenge.} 
\cjt is highly suitable for the interactive dashboard: Offline, we build \cjt for \dashq . Online, given \intq , \cjt ensures that computation needed is proportional to {\it the difference between it and \dashq} 
(Steiner tree);
such a difference is generally small, thanks to the incremental nature of the dashboard interactions.
However, one challenge arises when users submit multiple \intqs, each building upon the previous one. As the number of iterations increases, the Steiner tree is likely to expand due to the growing differences.
Ideally, we would want to calibrate not only \dashq but also \intqs. However, \intqs are only available online; calibrating them will slow down user interactions.
In the next section, we present an optimization to hide such a slowdown. The insight is that users typically have "think-times"~\cite{eichmann2020idebench} during interactions; we leverage it 
to calibrate \intqs in the background  without affecting users.

\section{System Overview}
\label{s:apps}

In this section, we provide the overview of \sys, a dashboard accelerator that manages \cjt to support interactive queries over join. We discuss \sys's usage,  architecture, and optimizations.

\begin{figure}
  \centering
  \includegraphics[width=0.8\columnwidth]{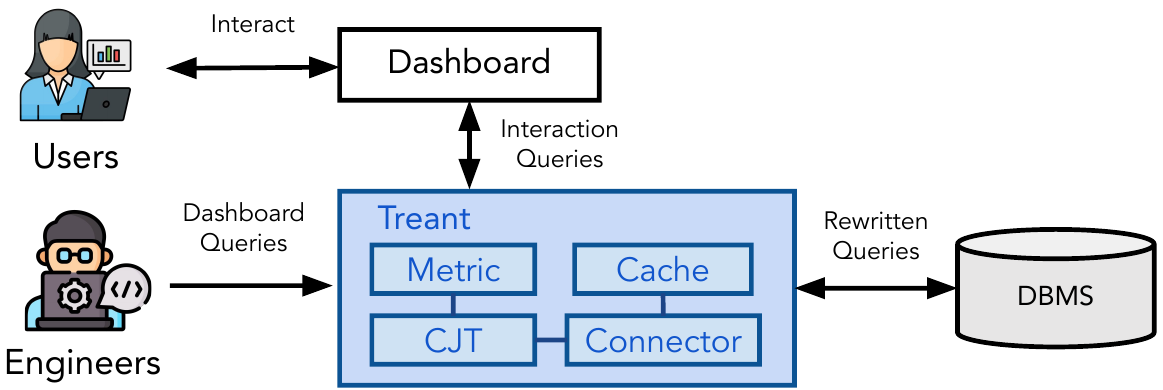}
   \vspace*{-3mm}
  \caption{\sys architecture.}
  \vspace*{-3mm}
  \label{fig:arch}
\end{figure}

\begin{figure}
  \centering
  \includegraphics[width=0.9\columnwidth]{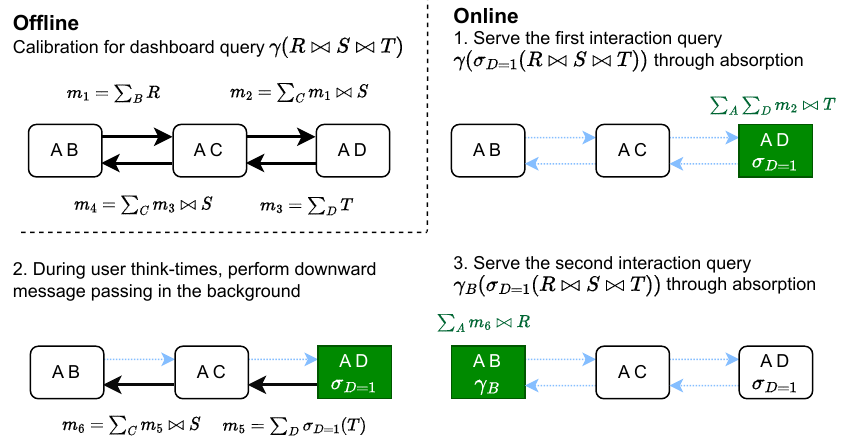}
  \vspace*{-5mm}
    \caption{Example \cjt management by \sys.}
    \vspace*{-3mm}
  \label{fig:workflow}
\end{figure}

\subsection{Usage Walkthrough}

We describe the detailed process of using \sys to build and use an interactive dashboard. \sys has both offline and online stages.

\subsubsection{Offline Stage.} The engineering team gathers data and constructs the dashboard offline through the following steps:

\stitle{Define Metrics.} Different domain users have different metrics of interest. For example, the sales department is interested in revenue,  while the marketing department is interested in return on investment (ROI).
The engineering team defines them as semi-ring aggregation (\Cref{s:background}) which express a wide range of aggregations.

\stitle{Construct Join Graph.} In enterprise data warehouses, there are typically a large number of tables for
metrics and enrichment dimensions  (join). The engineering team constructs a join graph that specifies these tables and the join conditions.

\stitle{Build Visualization.} The engineering team specifies the visualizations in the dashboard. Each visualization encodes data from a \dashq. For instance, a bar chart encodes a \dashq with one group-by, while a heatmap encodes one with two group-bys. \sys provides basic visualizations for the dashboard, but is also compatible with any external visualization system.

Finally, \sys takes as input the \dashqs (with metrics as semi-ring aggregations and join graphs in join clauses), connects to DBMS and pre-processes them for future dashboard interactions.

\subsubsection{Online Stage.} Domain users navigate dashboard to analyze metrics of interest. They interact with the dashboard through widgets, which in turn trigger \intqs that modify the initial \dashq. For example, a drop-down menu can modify the group-by attribute, while a slider can change the selected month of the \dashq.
\sys supports \intqs that:

\begin{myitemize}
  \item modify select/group clause  of the \dashq  (\Cref{sec:msgpassing})
  \item update or remove table in the join clause (\Cref{sec:msgpassing})
  \item join with new table that create/affect one bag (\Cref{sec:data_aug})
\end{myitemize}

\subsection{Architecture}

We first describe how \sys manages \cjt for the interactive dashboard, and then delve into the internal components.

\subsubsection{Management of \cjts.}
\label{sec:cjtmanage}
\sys manages \cjts for work sharing between \dashqs and \intqs.
\sys first builds \cjts for \dashqs offline to accelerate \intq online. However, users are likely to engage in more interactions that incrementally modify the previous \intq. \sys further calibrates \intq online: for each visualization (\dashq) and user session, \sys builds \cjt for the latest \intq during users' "think-times"~\cite{eichmann2020idebench} in the background.
Note that the calibration {\it doesn't need to be complete} and is halted  on receiving the next \intq  to not degrade interactivity. \sys can use the partially finished \cjt and take advantage of the finished messages to speed up \intq.

\begin{example}
 We illustrate the \cjt management using the example join graph (\Cref{fig:junctionHypertree}) in \Cref{fig:workflow}. Given a visualization of a single number with \dashq $Q_1=\gamma(R\Join S\Join T)$, \sys builds its \cjt offline. During online phase, user interact with the dashboard with  \intq $Q_2=\gamma(\sigma_{D_1}(R\Join S\Join T))$, and \sys uses $Q_1$'s \cjt to share messages. During user think-times, \sys calibrates $Q_2$ in the background. User performs the next interaction with  \intq $Q_3=\gamma_B(\sigma_{D_1}(R\Join S\Join T))$, and \sys uses $Q_2$'s \cjt. 
\end{example}

\subsubsection{Internal Components.}
\label{sec:internal}\sys internals are shown in \Cref{fig:arch}. In contrast to previous factorized systems~\cite{khamis2018ac,schleich2019layered} that use custom engines, \sys is a middleware that sits between the dashboard and users' DBMSes. It takes a \dashq or \intq as input, applies {\it pure query rewriting},
uses \cjt to determine the necessary messages to be computed, and computes messages by issuing SPJA queries to DBMSes. This makes \sys portable to any DBMS that executes SPJA queries.
The contents of messages (SPJA query results)
are 
stored as tables in the DBMS and \sys stores the pointers (table names) to these messages.

Offline, \sys establishes a connection to DBMS through {\it connectors}. For each visualization, the data engineer specifies its \dashq (with semi-ring aggregation) and \sys stores it in the {\it metric} component. Then, \sys re-writes \dashq for message passing and builds \cjt to pre-compute messages.

Online, \sys takes an \intq as input, finds the corresponding \cjt of previous \intq (\Cref{{sec:cjtmanage}}) based on the user session and the visualization, if available, or uses the \dashq otherwise. Then \sys uses that \cjt to pass messages only within the Steiner tree (\Cref{sss:cjtqexec}), performs absorption, retrieves the results and sends them to the dashboard for visualizations. 
The created messages are similarly stored within the DBMS, and only the pointers are returned. 
In the background, \sys further calibrates \intq (\Cref{{sec:cjtmanage}})

\revise{Finally, while the focus of \sys is to share messages from the most recent \cjt (the current dashboard state) for single-user session, there can be work-sharing opportunities  from (1) prior partially calibrated \jt and (2) \cjt across user sessions.
At present, \sys takes advantage of these opportunities through a message-level {\it cache}: For each message, \sys encodes its query definitions and the upstream sub-tree of the messages as the cache key (adequate to uniquely identify the message as per \Cref{prop:messagereuse}), with the message pointer as value. Before sending a message query to the DBMS, \sys checks if the message is cached.
The cache utilizes an LRU replacement policy by default, but refrains from removing messages if they are referenced by \cjts from \dashq or active user sessions. 
Upon removal of a message from the cache, \sys submits a deletion query for the message to the DBMS.}.

\begin{figure}
  \centering
  \includegraphics[width=.8\columnwidth]{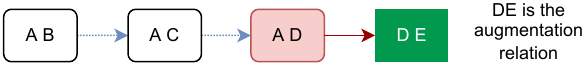}
   \vspace*{-3mm}
  \caption{Augmenting the join graph with \texttt{DE}. The Steiner tree is  \texttt{AD} $\rightarrow$ \texttt{DE} with root \texttt{DE} and requires one message (red).}
  \vspace*{-3mm}
  \label{fig:augmentation}
\end{figure}

\subsection{Augmentation Optimization}
\label{sec:data_aug}

Simple ML models like linear regression are widely used to examine  attribute relationships in the dashboard~\cite{morton2012trends}.
Data and feature augmentation~\cite{chepurko2020arda} further identify datasets to join with an existing training corpus in order to provide more informative features, 
and is a promising application on top of data warehouses and markets~\cite{fernandez2020data,chepurko2020arda,fernandez2018aurum}.   However, the major bottleneck is the cost of joining each augmentation dataset and then retraining the ML model.

The SOTA factorized ML~\cite{nikolic2018incremental,schleich2016learning,schleich2019layered,curtin2020rk} avoids join materialization when training models over join graphs.  First, it designs semi-ring structures for common models (linear regression~\cite{schleich2016learning}, factorization machines~\cite{schleich2021structure}, k-means~\cite{curtin2020rk}), and then performs 
upward message passing through the join graph.    If we augment with relation $r$, then factorized learning approaches execute the message passing through the whole augmented join graph again.  

In contrast, \cjt allows us to choose any bag $b$ that contains the join keys, construct an edge $b\to r$, and perform message passing using $r$ as the root.  In this setting, the Steiner tree is exactly 2 bags, and the rest of the messages in the \cjt can be reused.    For instance, \Cref{fig:augmentation} shows a join graph $\texttt{AB}\to \texttt{AC}\to \texttt{AD}$ that we augment with \texttt{DE}.  The Steiner tree is simply \texttt{AD} and \texttt{DE}, and we only need to send one message to compute the updated ML model.

\begin{figure}[t]
  \centering
  \begin{subfigure}{0.19\textwidth}
         \centering
         \includegraphics[width=\textwidth]{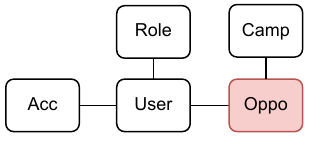}
         \vspace*{-5mm}
         \caption{\textbf{Salesforce}}
         \vspace*{-4mm}
\end{subfigure}
\hfill
\begin{subfigure}{0.24\textwidth}
         \centering
         \includegraphics[width=\textwidth]{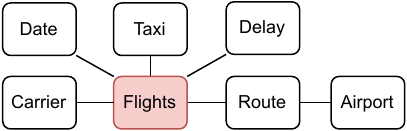}
         \vspace*{-4mm}
         \caption{\textbf{Flight}}
         \vspace*{-4mm}
\end{subfigure}
  \caption{Database schema. \red{Red} relation is the largest.}
  \vspace*{-3mm}
   \label{salesforce_schema}
\end{figure}

\vspace{-2mm}
\subsection{Limitations and Future Works}
\label{sec:limit}

\revise{The focus of this paper is on the message sharing opportunities, demonstrated through a simplified class of SPJA queries. There are limitations in the types of queries currently supported, and we aim to enhance expressiveness in future work to broaden applicability.}.

\vspace{-1mm}
\stitle{Limitations.} \revise{The current implementation uses parameterized SPJA based on exact matches, as per the format in \Cref{sec:msgpassing} and does not parse general SQL strings that can be rewritten to match this format.
In terms of expressiveness, \sys supports predicates over only attributes in a single bag for selection; for a predicate clause that references attributes from different
bags (e.g., $T.a = R.b$), \sys treats it as a post-processing of group-bys, which could be inefficient. Selection over aggregate query results is unsupported.
For aggregation, \sys only supports semi-ring aggregations; while semi-ring can express almost all commonly used  aggregations, percentile-based (e.g., median), and distinct-based (e.g., distinct count) aggregates are not supported.
 As for augmentation, \sys currently allows changes that create or affect a single bag; augmentations with attributes spanning across multiple bags are not permitted. Beyond SPJA queries, \sys supports \texttt{ORDER BY} and \texttt{LIMIT}, but as a naive post-processing over the SPJA query result that's not optimized.}.
 
\vspace{-1mm}
\stitle{Future works.} \revise{We aim to support semantically equivalent rewrites using existing methods like SPES~\cite{zhou2019automated}. For selections that reference attributes spanning multiple bags, recent optimization~\cite{khamis2020functional} can be applied to build range searching data structure for inequality predicates. For predicate that references results from other SPJA queries, a hierarchical \cjt~\cite{wu2007hierarchical} that references other \cjts  is a prospective direction. 
For percentile-based aggregations, we plan to apply semi-ring approximations~\cite{cormode2011sketch}.
Dealing with augmentations where join keys span multiple bags is another challenge we hope to address, as discussed in \Cref{s:augmentation}. For \texttt{ORDER BY} and \texttt{LIMIT}, we plan to integrate TOP-N optimizations~\cite{donjerkovic1999probabilistic} and approximations from MAP~\cite{wainwright2005map,conaty2017approximation} with \sys.}.

\section{Experiments}

Can \sys support \intqs at interactive speeds? What is the overhead? How well can \sys handle more complex applications like ML augmentation? 
We conducted experiments on both a single node DBMS (DuckDB~\cite{raasveldt2019duckdb}) and a cloud DBMS (Redshift).

\subsection{Single-node DBMS Experiments}
\label{s:local_exp}

\revise{
Our single-node DBMS experiments are designed to emulate the settings of local processing on a laptop.
We evaluate \sys on DuckDB~\cite{raasveldt2019duckdb} due to its popularity: DuckDB is an OLAP DBMS with  superior single-node performance and seamless integration with other Python/R data analytics libraries.}.

\stitle{Setup.}
\revise{We use two datasets for dashboards: \textbf{Salesforce}\cite{salesforce}, a public dataset provided by Sigma Computing for CRM and marketing analysis with 36 numerical and 229 categorical attributes; and \textbf{Flight}\cite{flight}, a real-world dataset with 27 numerical and 5 categorical attributes, commonly used in interactive data exploration~\cite{eichmann2020idebench}. 
However, \textbf{Salesforce} is a demo dataset of only $27MB$, while modern laptop can process $>10 GB$ data with DuckDB.}.
To address this discrepancy, we employ the data scaler in  IDEBench~\cite{abourezq2016database} to scale both datasets. The scaling process involves denormalization, estimating the distribution, sampling rows from the distribution, and finally normalizing the table through vertical partitioning.
\revise{We scale \textbf{Salesforce} to 50M rows for 13.7GB, and \textbf{Flight} for 300M rows for 15GB.}.
The final normalized schemas are in \Cref{salesforce_schema}.

\begin{figure}
  \centering
  \includegraphics[width=.5\columnwidth]{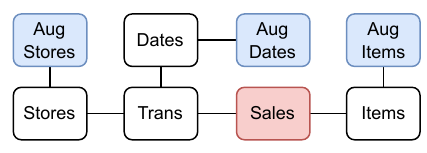}
  \vspace*{-3mm}
  \caption{\textbf{Favorita} schema. \textbf{\red{Sales}} is the largest relation. \textbf{\blue{Aug Stores}}, \textbf{\blue{Aug Dates}} and \textbf{\blue{Aug Items}} are augmentation relations.}
  \vspace*{-3mm}
   \label{fav_schema}
\end{figure}

For ML augmentation, we use the \textbf{Favorita}~\cite{favorita} dataset of purchasing and sales forecasts, widely used in prior factorized ML~\cite{schleich2016learning, schleich2019layered} (see \Cref{fav_schema} for the schema). \textbf{Sales} is the largest relation (241MB), while the others are $<2MB$. \revise{All experiments were run on GCP c2d-standard-4 (4 vCPUs, 16 GB RAM).}.

\subsubsection{Salesforce Dashboard}
\label{sec:singledashboard}
We conduct experiments on the Salesforce dashboard~\cite{salesforce} built by Sigma Computing.

\stitle{Workloads.} The Salesforce dashboard tracks various metrics such as pipeline, and productivity, which are all computed using the sum aggregations. However, we find that the specific metric chosen has little impact on the query performance. Therefore, we randomly choose the total pipeline amounts as the metric.

We consider two types of visualizations from the Salesforce dashboard illustrated in \Cref{fig:sigmadashboard}: a single value visualization, which corresponds to a \dashq without group-by or selection, and a pie chart that is grouped by Campaign (\textbf{Camp}) type. The dashboard includes drop-down lists for selecting the user name, title, campaign start date, role name, and group-by user title or account state; we experiment with all of these interactions. \revise{We further conduct tests of interactions that modify the \textbf{Camp} relation (by random cell value perturbations) and remove the \textbf{Acc} relation.}.

\stitle{Baselines.} We consider the following baselines

\begin{figure}
  \centering
  \begin{subfigure}[c]{0.5\textwidth}
     \centering
     \includegraphics[width=\textwidth]{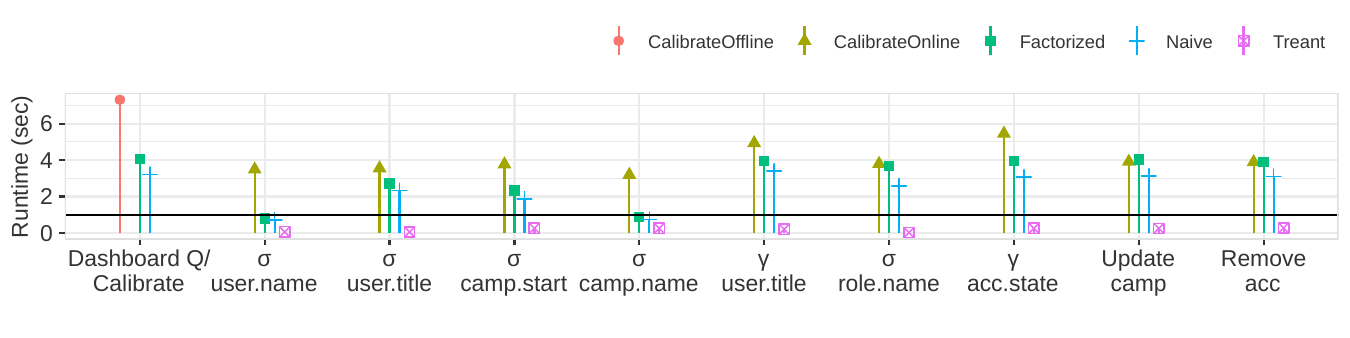}
     \vspace*{-8mm}
     \caption{Performance for  single value visualization.}
     \label{fig:salesforce1}
 \end{subfigure}
 \hfill
  \begin{subfigure}[c]{0.5\textwidth}
     \centering
     \includegraphics[width=\textwidth]{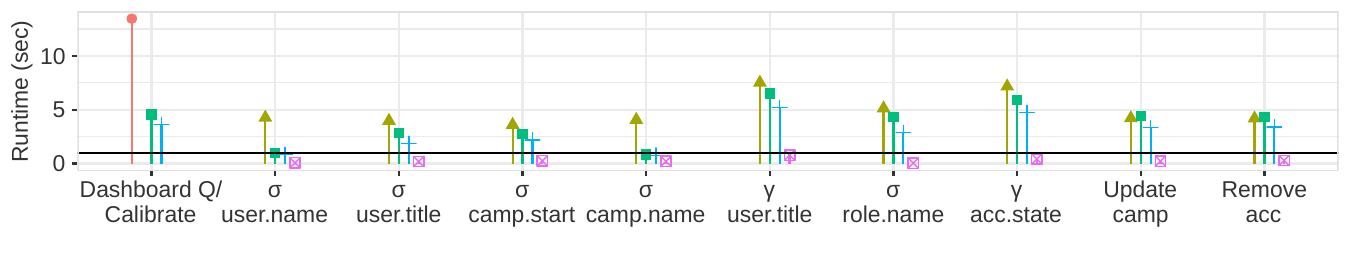}
     \vspace*{-8mm}
     \caption{Performance for  pie chart grouped by Campaign Type.}
     \label{fig:salesforce2}
 \end{subfigure}
 \hfill
\vspace*{-3mm}
  \caption{\textbf{Salesforce} Dashboard Performance. Horizontal line shows the $1s$ interactive response threshold.}
  \vspace*{-3mm}
  \label{fig:salesforce}
\end{figure}

\begin{myitemize}
  \item {\bf Naive}: Execute the naive SPJA queries translated from dashboard interactions in the DBMS without implementing factorized query execution nor work-sharing from pre-computed data structures.
  \item {\bf Factorized}: Rewrite the naive queries into message passing for factorized execution, but still doesn't exploit work-sharing.
   \revise{\item \sys  consists of multiple stages for online \intq execution (directly experienced by the users)  and the offline/background  calibration overheads. We report them separately:
\begin{myitemize}
\item {\bf Treant}: online \intq execution using \cjts (from prior \intq or \dashq) for acceleration.
\item {\bf CalibrateOffline}: offline calibration overhead to construct \cjts for the initial \dashqs.
\item {\bf CalibrateOnline}: background calibration overhead to construct \cjt for the current \intq (to proactively expedite  future queries), which doesn't need full completion.
\end{myitemize}}.
\end{myitemize}
\vspace*{-1mm}
\stitle{Results.} 
\Cref{fig:salesforce} presents the results. 
The performance of both \textbf{Factorized} and \textbf{Naive} varies depending on the type of interaction, with faster performance for selection due to smaller data sizes to process.  However, they still take $> 2s$ for most \intqs and can be as slow as $7s$.
We find that
factorized execution (\textbf{Factorized}) alone results in even slower performance than \textbf{Naive}.  This is because \textbf{Factorized} is optimized for many-to-many joins, which are not present in \textbf{Salesforce} workloads. Additionally,  \textbf{Factorized} introduces additional aggregations for each join edge.

In contrast, \textbf{Treant} is able to execute various types of \intqs, from selection, group-by, to relation update and removal, within $100ms$  by reusing messages, offering two orders of magnitude improvements; the  offline overhead  (\textbf{CalibrationOffline}) is only ${\sim}2\times$ the cost of executing  the \dashq by \textbf{Factorized}. To ensure quick responses in future interactions,  \sys calibrates \intq (\textbf{CalibrationOnline}) whose time is at the same scale as \textbf{Factorized} as it only requires downward message passing (\Cref{sec:calibration}), and is well within user think-times (${<}10s$~\cite{eichmann2020idebench}). Furthermore, \textbf{CalibrationOnline} is in the background during user think-times, and doesn't require full completion. \revise{Regarding storage, the intermediate messages only occupy  $363 MB$ ($<3\%$ of DB size).}.
\vspace*{-1mm}

\subsubsection{Flight Dashboard}
\label{sec:flight}
We next experiment with the Flight dataset~\cite{flight}.

\stitle{Workloads.} IDEBench~\cite{eichmann2020idebench} produces a random workload consisting of a collection of visualizations (\dashqs) and, for each visualization, a series of \intqs that progressively incorporate selections.
We use the default workload\footnote{\texttt{independent} in \url{https://github.com/IDEBench/IDEBench-public/tree/master/data}}, which contains 8 total \intqs across 5 visualizations.  We use the same baselines as in \Cref{sec:singledashboard}. However, to demonstrate the advantage of online calibration, we evaluate {\bf Tre+Offline}, which only uses \cjts created offline. Based on IDEBench recommendation, we use a think-time of $10$s by default, and study the sensitivity later.

\begin{figure}
  \centering
  \begin{subfigure}[c]{0.35
  \textwidth}
     \centering
     \includegraphics[width=0.8\textwidth]{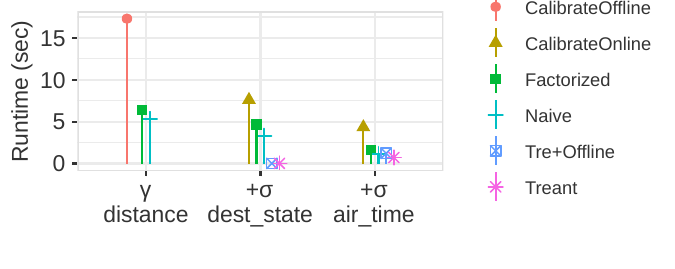}
     \vspace*{-3mm}
 \end{subfigure}
 \begin{subfigure}[c]{0.21\textwidth}
     \centering
     \includegraphics[width=0.8\textwidth]{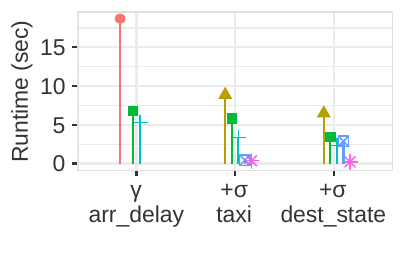}
     \vspace*{-3mm}
 \end{subfigure}
  \begin{subfigure}[c]{0.21\textwidth}
     \centering
     \includegraphics[width=0.8\textwidth]{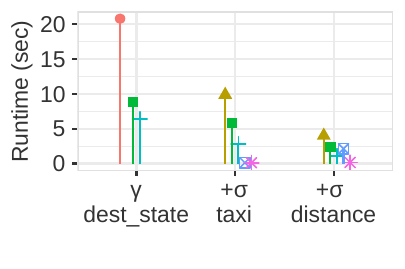}
     \vspace*{-3mm}
 \end{subfigure}
  \begin{subfigure}[c]{0.21\textwidth}
     \centering
     \includegraphics[width=0.8\textwidth]{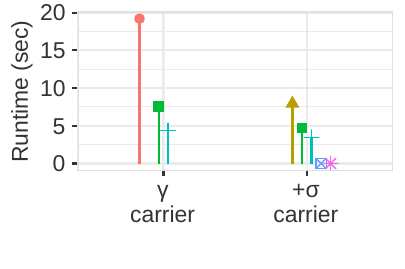}
     \vspace*{-3mm}
 \end{subfigure}
  \begin{subfigure}[c]{0.21\textwidth}
     \centering
     \includegraphics[width=0.8\textwidth]{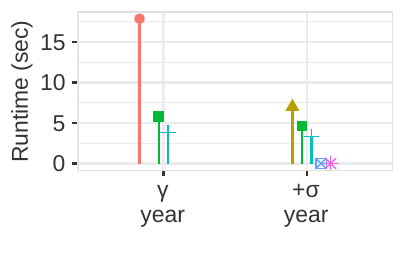}
     \vspace*{-3mm}
 \end{subfigure}
\vspace*{-3mm}
  \caption{\textbf{Flights} Dashboard Performance. The dashboard features five visualizations. For each visualization, the leftmost is \dashq. \intqs progressively adds ($+$) selection or group-by element to the previous one.}
  \vspace*{-5mm}
  \label{fig:ide}
\end{figure}

\begin{figure}
  \centering
  \includegraphics[width=.6\columnwidth]{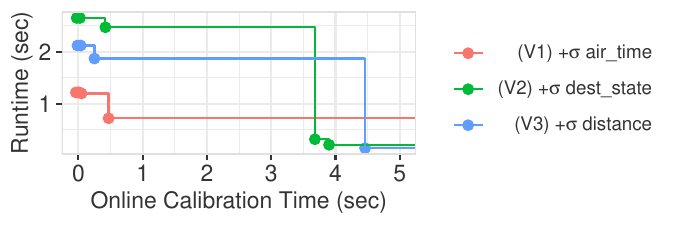}
  \vspace*{-5mm}
  \caption{\revise{Runtime for the second \intq over the first three visualizations, with varying online calibration.}.}
  \vspace*{-3mm}
   \label{fig:think}
\end{figure}

\begin{figure}
  \centering
  \begin{subfigure}[c]{0.5\textwidth}
     \centering
     \includegraphics[width=0.8\textwidth]{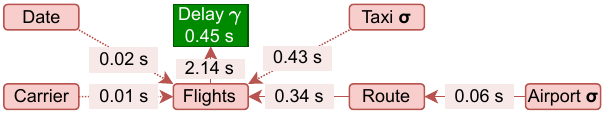}
     \vspace*{-2mm}
     \caption{\textbf{Factorized} performs message passing without sharing.}
 \end{subfigure}
 \begin{subfigure}[c]{0.5\textwidth}
     \centering
     \includegraphics[width=0.8\textwidth]{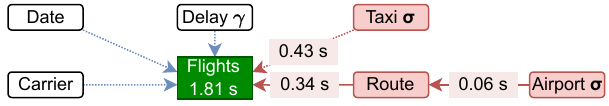}
     \vspace*{-2mm}
     \caption{{\bf Tre+Offline} shares messages, but only  with offline \cjt.}
 \end{subfigure}
 \begin{subfigure}[c]{0.5\textwidth}
     \centering
     \includegraphics[width=0.78\textwidth]{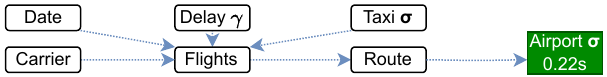}
     \vspace*{-2mm}
     \caption{\sys enhances message sharing via online calibration.}
 \end{subfigure}
  \vspace*{-4mm}
  \caption{\revise{Case study for the second \intq over the second visualization. Blue dotted lines represent shared messages, while red lines indicate computed messages with respective runtimes noted on the edges. The absorption runtimes are displayed in the green rectangle bag.}. }
  \vspace*{-3mm}
   \label{fig:case}
\end{figure}

\stitle{Results.}
The results are displayed in \Cref{fig:ide}. Both \textbf{Factorized} and \textbf{Naive} take $3{-}6s$ for most queries. For \sys, offline calibration is ${\sim}3\times$ \textbf{Factorized} due to the larger message size during the group-bys, but reduces \textbf{Treant} to ${<}200ms$.   Online calibration takes $4-9s$, well within the think-time of $10s$. 
For the second \intqs of the $2^{nd}$ and $3^{rd}$ visualizations, using only offline-created \cjt ({\bf Tre+Offline}) takes $>2s$ due to the larger Steiner tree size, and online calibration reduces this time by ${>}10\times$. \revise{In terms of storage, the intermediate messages take up just $89 MB$ ($<1\%$ of DB size).}.

\stitle{Online Calibration Sensitivity.}
\revise{For online calibration, think-time may not always be possible to leverage. For instance, in multi-tenant DBMS, other users may execute their queries concurrently while one user is thinking. To examine this, we study the second \intq in the first three visualizations, whose runtime depends on the online calibration. \Cref{fig:think} presents the \intq  runtime with varying online calibration time. The plot shows a stepped pattern, as reductions in \intq runtime only occur upon completion of sharable message computations. Consequently, each step corresponds to a completed sharable message. Without online calibration, the runtime equals that of {\bf Tre+Offline}. As online calibration time increases, the runtime decreases accordingly due to partial calibration, and reaches ${<}1s$ after only ${<}5s$.}.

\stitle{Case Study.} 
\revise{To understand where the performance improvements stem from, we conduct a case study on the second \intq over the second visualization, because it shows the most significant improvement. \Cref{fig:case} shows the detailed runtime for each message and absorption  for different baselines. \textbf{Factorized}  performs message passing without sharing, while {\bf Tre+Offline} shares messages but only with the offline \cjt. They are slowed  by the costly message passing or absorption over the large fact table  (${>}1.8s$). 
In contrast, {\bf Treant} greatly improves message sharing using online calibration. Once the online calibration is completed, the \jt is calibrated. For the interaction that selects Airport attribute, this only requires an absorption over Airport (small dimension table) in $0.22s$.}.

\subsubsection{ML augmentation}
\label{sec:mlaugexp}
We next evaluate the benefit of \sys for ML augmentation using the \textbf{Favorita}  dataset.

\stitle{Workloads.}  We train linear regression using  (\texttt{Sales}.unit\_sales, \\\texttt{Stores}.type, \texttt{Items}.perishable) as features, and \texttt{Trans}.transactions (number of transactions per store, date) as the target variable $Y$. 

To simulate a data warehouse with augmentation data of varying effectiveness, we generate synthetic data to augment (join) with \texttt{Dates}, \texttt{Stores}, and \texttt{Items}.  For each of these three relations, we first generate a predictive feature $\hat{Y}$ as the average of Y grouped by primary key.  We then create 10 augmentation relations with schema \texttt{(k,v)}, where \texttt{k} is the primary key and \texttt{v} varies in correlation $\hat{Y}$~\cite{kaiser1962sample}:  The correlation coefficient $\phi$ is drawn from the inverse exponential distribution $min(1, 1/Exp(10))$, and the values are the weighed average between $\hat{Y}$ and a random variable weighed by $\phi$. We individually evaluate the model accuracy ($R2$) for each of the 30 augmentation relations, and measure the cumulative runtimes. 

In addition to {\bf Treant}, we  also compare the training time of \textbf{Fac} that applies factorized ML but trains each model independently without work sharing, and \textbf{LMFAO}~\cite{schleich2019layered}, the SOTA factorized ML system that is algorithmically similar to \textbf{Fac} but implemented with a custom engine in C++ and not portable to user DBMSes.  To ensure a fair comparison, we exclude the time required to read files from disk and the compilation time for \textbf{LMFAO}, but include all the time needed to build data structures and run queries.

\begin{figure}
  \begin{subfigure}[t]{0.49\columnwidth}
  \centering
    \includegraphics[width=0.9\textwidth]{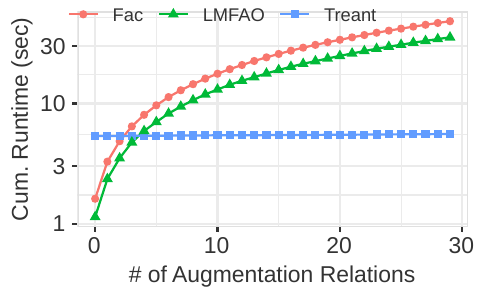}
    \vspace*{-3mm}
    \caption{Cumulative runtime (log)}
     \label{exp:aug_time}
     \end{subfigure}
     \hfill
    \begin{subfigure}[t]{0.49\columnwidth}
    \centering
    \includegraphics[width=0.9\textwidth]{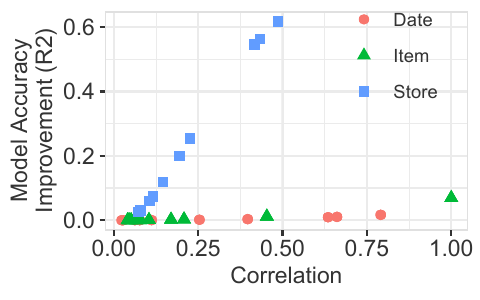}
    \vspace*{-3mm}
    \caption{Accuracy (R2) improvements.}
    \label{exp:aug_r2}
     \end{subfigure}
     \vspace*{-3mm}
  \caption{Augmentation run time and model performance.}
  \vspace*{-3mm}
\end{figure}

\stitle{Results.} 
\Cref{exp:aug_time} reports the cumulative runtime to augment and retrain the model.
\textbf{Fac} takes ${>}1.3$ min, while {\bf Treant} takes ${\sim}6s$: calibration dominates the cost, and is ${\sim}2\times$ the cost of training a single model because of the downward message passing. However, after calibration, {\bf Treant}  evaluates all 30 augmentations in ${<}1s$.  
\textbf{LMFAO} takes ${\sim}1.3\times$ less time than \textbf{Fac} for model training due to implementation difference, but even when including the offline calibration cost,  {\bf Treant} is ${\sim}13\times$ faster than \textbf{LMFAO} after 30 augmentations.
\Cref{exp:aug_r2} reports the accuracy improvement above the baseline ($0.031$) after each augmentation, and we see a wide discrepancy between good and bad augmentations ($+0$ to $+0.61$).

\subsection{Cloud DBMS Experiments}
\label{exp:cloud}
We now evaluate  \sys on the cloud DBMS (AWS Redshift).

\stitle{Setup.} 
We use TPC-H (SF=50) for dashboard and TPC-DS (\revise{SF=50}.) for empty bag optimizations.
We used dc2.large node (2 vCPU, 15GB memory, 0.16TB SSD, 0.60 GB/s I/O). All experiments warm the cache by pre-executing queries until the runtime stabilizes.

\begin{figure}
  \centering
      \includegraphics [width=0.9\columnwidth] {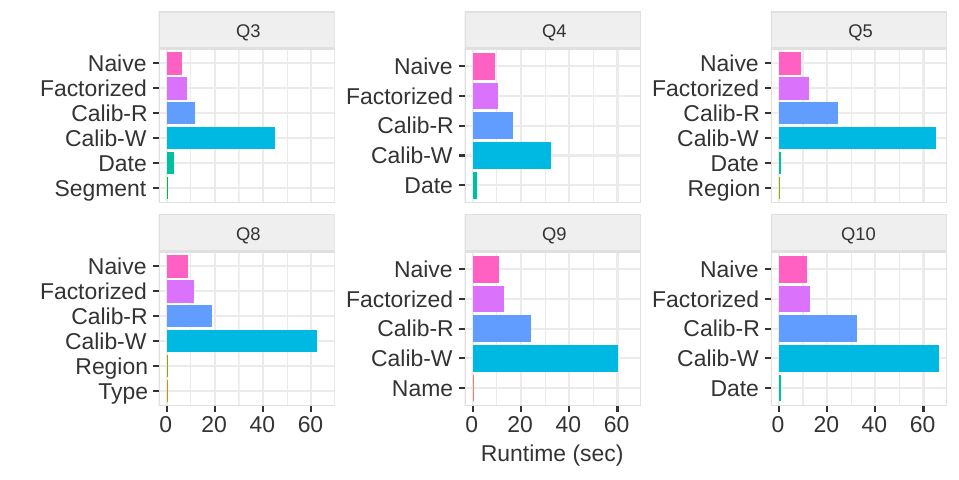}
      \vspace*{-5mm}
  \caption{Run time for TPC-H dashboard. \textbf{Naive} executes queries without message passing. \textbf{Factorized} executes queries with message passing. \textbf{Calib-R} computes messages for calibration without materializing them, and \textbf{Calib-W} materializes them. The remaining bars are for \intqs that vary the values of the parameters (labels).}
  \vspace*{-5mm}
  \label{fig:tpch_runtime}
\end{figure}

\begin{figure}
  \centering
  \begin{subfigure}[c]{0.25\textwidth}
     \centering
     \includegraphics[width=0.8\textwidth]{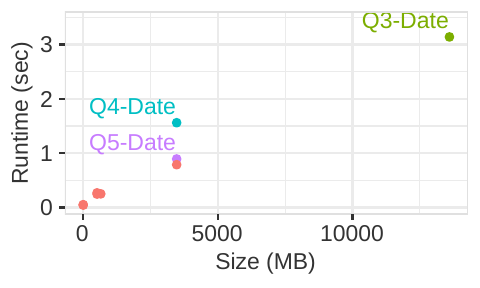}
     \vspace*{-3mm}
     \caption{}
     \vspace*{-3mm}
     \label{fig:tpch_runtime_size}
 \end{subfigure}
 \hfill
  \begin{subfigure}[c]{0.18\textwidth}
     \centering
     \includegraphics[width=0.8\textwidth]{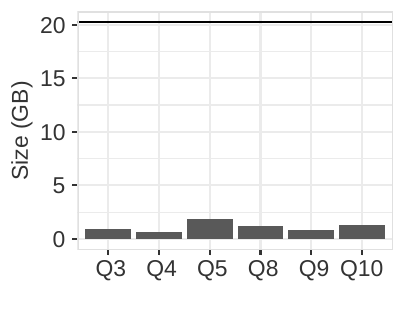}
     \vspace*{-3mm}
     \caption{}
     \vspace*{-3mm}
     \label{fig:tpch_size}
 \end{subfigure}
 \hfill
  \caption{(a) Size of annotated bag (by predicate) vs query runtime. (b) Total message size overhead of \sys. Horizontal line is the TPC-H database size (${\sim}20$ GB).}
\end{figure}

\subsubsection{Interactive Dashboard}
\label{sec:tpch}
We evaluate \sys on TPC-H queries.

\stitle{Workloads.} We build an interactive dashboard based on a subset of the TPC-H queries (Q3-5,8-10) that can be rewritten as SPJA queries (see \Cref{s:tpchqueries}). These TPC-H queries are parameterized, so we construct a \dashq for each using random parameter values and then create \intqs that vary each parameter.

We compared the runtime of the \intq using different approaches: \textbf{Naive} simply executes the query on Redshift; \textbf{Factorized}  rewrites the query as message passing for factorized query execution; and \sys. For \sys, we reported calibration execution cost (\textbf{Calib-R}) separately from the calibration materialization cost (\textbf{Calib-W}), since writes on Redshift are particularly expensive.

\stitle{Results.} \Cref{fig:tpch_runtime} shows the run time. 
Calibration (\textbf{Calib-W}) takes $4{\sim}7\times$ longer than \textbf{Naive}.  As expected, upward and downward message passing alone is ${\sim}2\times$ slower (\textbf{Calib-R}), and the rest is dominated by high write overheads;   Q8 groups by 2 attributes, so its message sizes are ${\sim}2\times$ larger, and $4\times$ slower overall. As in \Cref{sec:singledashboard}, \textbf{Factorized} is slower than \textbf{Naive} because there is no many-to-many join and it has additional aggregation overheads.

In contrast, \sys accelerates TPC-H queries by nearly $1000\times$ over \textbf{Naive} for parameters including Segment, Region and Type.  Naturally, the speedup depends linearly on the size of the bag that contains the parameterized attribute (\Cref{fig:tpch_runtime_size}). Q3 Date incurs a higher cost as it includes the fact table in Steiner tree, which could be optimized by creating an empty bag for Date (\Cref{sec:jtdatastructure}). We note that the space overhead for calibration is only ${<}2GB$ compared to the original database size ${\sim}20 GB$ (\Cref{fig:tpch_size}). This is because messages are aggregated results from these relations.

\subsubsection{Empty Bag Optimization}
\label{sec:emptyexp}
Empty bags are a novel extension to materialize custom views. We evaluate the costs and benefits of empty bags using TPC-DS. We create an empty bag \textbf{(Store,Time)} as illustrated in \Cref{fig:tpc_ds_empty}. Then, we query the  maximum count of sales for all stores and times: ${\bf Q} = \gamma_{MAX(COUNT)}(\gamma_{COUNT(\cdot), Store, Time}(\Join))$ in two unique ways: (1). Without Empty Bag, ${\bf Q}$ is executed by first aggregating the count over the absorption result of \textbf{Store\_Sales}, since \textbf{Store\_Sales} is the only bag contains both \textbf{Store} and \textbf{Time} dimensions, then computing the max sales. (2). With Empty Bag, ${\bf Q}$ is executed directly over the absorption result of the empty bag, which is sufficient to answer aggregation queries over \textbf{(Store,Time)}.

\begin{figure}
\centering
      \includegraphics [width=0.25\textwidth] {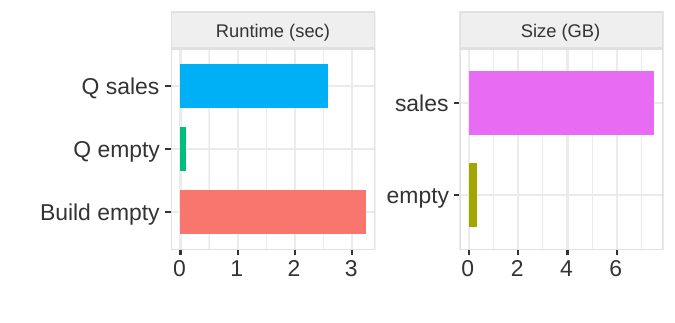}
      \vspace*{-5mm}
      \caption{Runtime to build and query the empty bag and \textbf{Store\_Sales} table, and their storage sizes.}
      \vspace*{-1mm}
      \label{exp:empty_bag}
\end{figure}

\revise{\Cref{exp:empty_bag} shows the runtimes and sizes. Empty bag takes ${\sim} 3s$ to build, and accelerates ${\bf Q}$ by ${\sim}25\times$. Additionally, the storage space required for the empty bag is $21\times$ smaller than that of \textbf{Store\_Sales}.}.

\vspace*{-2mm}
\section{Related Work}\label{s:related}
\vspace*{-2mm}
\stitle{Interactive Queries.} Previous  works use indexing and data cubes~\cite{gray1997data} to support interactive queries. Some studies have improved them to Nanocubes~\cite{lins2013nanocubes} for spatiotemporal data and Hashedcubes~\cite{pahins2016hashedcubes} with additional optimizations, or use sophisticated materialization techniques~\cite{liu2013immens,moritz2019falcon}. However, these approaches often have high preprocessing overhead (e.g., taking hours for 200MB data~\cite{lins2013nanocubes}), and require denormalization for large joins.
In contrast, \cjt has been shown to be exponentially more efficient for {\it interactive queries} over large joins (\Cref{app:olap}) than data cubes with a constant factor overhead.

\stitle{Early Marginalization.} Early Marginalization was first introduced by Gupta et al.~\cite{gupta1995aggregate} as a generalized projection for simple e.g., count, sum, max queries. It was extended by factorized databases to compactly store relational tables~\cite{olteanu2015size} and quickly execute semi-ring aggregation queries~\cite{schleich2019layered,joglekar2015aggregations}. Abo et al.~\cite{abo2016faq} generalize early marginalization and establish the equivalence between early marginalization and variable elimination in Probabilistic Graphical Models~\cite{koller2009probabilistic}. 
However, prior works~\cite{schleich2019layered} only share work within a query batch but not between batches for interactive queries.

\stitle{Calibrated Junction Tree.} Calibration Junction Tree was first proposed by Shafer and Shenoy~\cite{shafer1990probability} to compute inference over probabilistic graphical models. While it has found extensive applications across fields such as engineering~\cite{zhu2015junction,ramirez2009fault}, ML~\cite{braun2016lifted,deng2014large}, and medicine~\cite{pineda2015novel,lauritzen2003graphical}, its use has been limited to probabilistic tables (for sum of probabilities). 
\revise{The calibration process is reminiscent of Yannakakis's two-pass semi-join reduction~\cite{yannakakis1981algorithms}. However, Yannakakis's algorithm mainly aims to eliminate redundant tuples in individual relations as a pre-processing step for single query execution. Furthermore, it does not materialize messages and is restricted to the 0/1 semi-ring. 
In contrast, \cjt materializes messages for future reuse across queries.
In this work, we broaden the scope of \cjt to general semi-ring aggregation for SPJA queries.}.

\stitle{Semantic Caching.} \revise{To accelerate online queries, semantic caching~\cite{ren2003semantic,ahmad2020enhanced,chaudhuri1995optimizing,gupta1995adapting, asgharzadeh2009exact} caches previous SPJA query results and reuses them for later SPJA queries, by taking row/column containment, predicate overlaps~\cite{theodoratos2004constructing,guo1996satisfiability}, and constraints into account~\cite{gupta1995adapting, ren2003semantic,gryz1998query}. 
The analysis for identifying reuse opportunities has also been applied to identify materialized views to reuse~\cite{li2005formal}.
In contrast, \sys specifically exploits the semiring properties of the aggregation functions to 1) leverage factorized query execution via message passing, 2) identify partial aggregates (messages) to cache during query execution, and 3) identify the best plan that reuses the cached partial aggregates.  Like caching, \sys executes and caches online.  In addition, \sys proactively populates the cache in anticipation of incremental changes to the most recently executed query, leveraging user think-time.
\sys currently only shares "messages" with identical query definitions and leave other types of sharing (e.g., row/column containment, predicate overlaps) as future works.
}.

\stitle{Multi-query optimization.} \revise{Multi-query optimization (MQO)~\cite{roy2000efficient,hong2009rule,schleich2019layered} shares the state and computation of subexpressions across queries (e.g., sharing scans across queries). However, it centers on batches of queries known a priori. The optimization of interactive dashboards can be considered as an online version of multi-query optimization. \sys approaches this by applying a practical heuristic that interactive queries are incremental, and uses "messages" as the core unit to reuse. These ideas could be integrated into MQO.}.
\vspace*{-3mm}

\section{Conclusions}

We present \sys, a dashboard accelerator over joins. \sys uses factorized query execution for aggregation queries over large joins, and proactively materializes messages, the core intermediates during factorized query execution.
that can be shared across  \intqs. 
To effectively manage and reuse messages, we introduced the novel Calibrated Junction Hypertree (\cjt) data structure. \cjt uses annotations to support SPJA queries, applies calibration to materialize messages in both directions and computes the Steiner tree to assess the reusability
of messages.
We implement \sys to manage \cjt as middleware between DBMSes and dashboards.
Our experiments evaluate \sys on a range of datasets on both single node and cloud DBMSes, and we find that \sys accelerates dashboard interactions by two orders of magnitude.

\balance
\bibliographystyle{abbrv}
\bibliography{paper}

\clearpage

\appendix

\section{Message Passing Algorithm}
\label{messagepassingalg}
\begin{algorithm}
        \caption{Message Passing and Calibration Algorithm}
        \label{message_pass}
        \begin{algorithmic}[1]
            \State // Pass Message from bag u to v where $u,v \in V$\;
            \Function{PassMessage}{$((E,V), \mathcal{X}, \mathcal{Y})$, u, v}\;
                \State // All the neighbours\;
                \State $N(u) = \{c | c\to u\in\mathcal{E}\}$\;
                \State // All incoming messages from in-neighbours except v\;
                \State $M(u) = \{\mathcal{Y}(i\to u) | i\in N(u)\land i\not\eq v\}$\;
                \State // Compute and store message from u to v\;
                \State $\mathcal{Y}(u\to v) = \sum_{u-v\cap u} \Join \left(M(u) \cup \mathcal{X}^{-1}(u) \right) $\;
            \EndFunction
            \State \;
            \State // Upward Message Passing to root $r \in V$ \;
            \Function{Upward}{$((E,V), \mathcal{X}, \mathcal{Y})$, r}
                \ForAll{Bag $c \in V - r$ from leaves to root r bottom up}\;
                    \State p = parent of c\;
                    \State PassMessage($((E,V), \mathcal{X}, \mathcal{Y})$, c, p)\;
                \EndFor
            \EndFunction
            \State \;
            \State // Downward Message Passing from root $r \in V$ \;
            \Function{Downward}{$((E,V), \mathcal{X}, \mathcal{Y})$, r}
                \ForAll{Bag $p \in V$ from root r to leaves top down}\;
                    \ForAll{child bag c of p}\;
                        \State PassMessage($((E,V), \mathcal{X}, \mathcal{Y})$, p, c)\;
                    \EndFor
                \EndFor
            \EndFunction
            \State \;
            \State // Calibrate Junction Hypertree \;
            \Function{Calibration}{$((E,V), \mathcal{X}, \mathcal{Y})$}
                \State // choose a random bag as root\;
                \State $r \in V $ \;
                \State Upward($((E,V), \mathcal{X}, \mathcal{Y})$, r)\;
                \State Downward($((E,V), \mathcal{X}, \mathcal{Y})$, r)\;
            \EndFunction
            
        \end{algorithmic}
    \end{algorithm}

\section{Feature Augmentation with \cjt.}
\label{s:augmentation}

While feature augmentations over single join key are efficient, those over multiples multiple join keys are complex. We need to query the \cjt group-by all join keys, which might result in a large Steiner Tree and we need to re-design the \jt after augmentation. 

\stitle{Feature Augmentation over Multiple Bag.} For Feature Augmentation over multiple, we want to query the aggregation group-by the join keys from \cjt. This could be considered as a SPJA query with group-by annotations, and can be computed through Upward Message Passing in the Steiner tree. 

\stitle{Connect Augmentation Relation to \jt.} To connect augmentation relation to \jt where the join key is distributed over multiple bag, we have to add all the join key to the bags of Steiner tree, create an augmentation bag containing augmentation relation, and connect the augmentation bag to any of the bag in Steiner tree. Notice that the \jt with added attributes in the bags can be inefficient, and we may redesign \jt to find a better one.

\stitle{Optimize \cjt design.} To optimize \cjt for feature augmentation, we create empty bags for common join keys. Consider TPC-DS as an example, whose (simplified) join graph (also \jt) is shown in \Cref{fig:tpc_ds_schema}. We can cluster time and stores in an empty bag shown in \Cref{fig:tpc_ds_empty} to support efficient augmentation of spatio-temporal features. 

\section{Selection Pushdown.}
\label{s:selectionopt}

\begin{figure}
  \centering
         \includegraphics[width=0.3\textwidth]{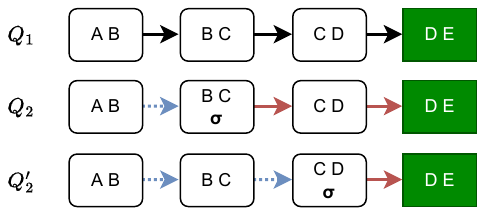}
         \caption{Message size vs reuse trade-off. Given total count query $Q_1$, $Q_2$ adds a predicate to C. Pushing down selection as in $Q_2$ may reduce message size but hinders reuse as compared to $Q'_2$. Dotted \blue{blue edges} are reusable messages and solid \red{red edges} are non-reusable edges.}
         \label{fig:selection_pushdown}
\end{figure}

\Cref{prop:messagereuse} implies that an annotation can ``block'' reuse along all of its downstream messages. For group-by annotation, we greedily push down it to the leaf of the connected subtree closest to the root to maximize reusability. 
However, pushing selections down trades-offs potentially smaller message sizes for limited reusability:

\begin{example}
  Suppose we have materialized messages for $Q_1$ in \Cref{fig:selection_pushdown}, and want to execute $Q_2$, which has an additional predicate over C.  If we annotated bag \texttt{BC} with $\sigma$, this may reduce the message size but we cannot reuse the message in $\texttt{BC} \to \texttt{CD}$. If we annotate \texttt{CD} ($Q'_2$), we can reuse the message but risk larger message sizes. 
\end{example}

In practice, we prioritize reuse by pulling annotations close to the root---reuse helps avoid scan, join, and aggregation costs, whereas larger message sizes simply increase scan sizes.

\section{Data Cube}
\label{app:olap}

OLAP data cubes~\cite{gray1997data} materialize a lattice of data cuboids parameterized by the set of attributes that future queries will filter/group by.  Traditionally, the data structure is built bottom-up in order to share computation---each cuboid is built by marginalizing out irrelevant attribute(s) from a descendant cuboid.  If the cube is over a join graph, then there is the additional cost of first materializing the (potentially very large) join result to compute the bottom cuboid.  Although prior work explored many optimizations (parallelization~\cite{wang2013scalable,dehne2002parallelizing,taniar2002parallel}, approximation~\cite{vitter1998data}, partial materialization~\cite{han1998selective,wang2002condensed}, early projection~\cite{kotsis2000elimination}), neither early marginalization nor work-sharing based on \cjts have been explored.

\cjts are a particularly good fit for building data cubes because, in practice, they are restricted to a small number of attributes in order to avoid exponentially large cuboids.    In this setting, we can build \cjts for a carefully selected set of pivot queries to accelerate cube construction by
 1) not materializing the full join graph when building the cuboids, and 2) aggressively reuse messages to answer OLAP queries not directly materialized by a cuboid.

\subsection{Complexity Analysis}
\label{app:cube_complexity}
Let us first analyze the complexity of using \cjts to answer OLAP queries.  This will provide the tools to trade-off between OLAP query performance and space requirements for materialization.

Let the database contain $r$ relations each with $O(n)$ rows, the domain of each attribute is $O(d)$, and the join graph contains $m$ unique attributes.  
Suppose we have calibrated each cuboid with $k$ group-by attributes (the pivot queries).  Calibration costs $O(rnd^k)$ for each pivot query where the cost per bag is $O(nd^k)$ (cross product between relation size $n$ and incoming messages). Since the output message size is also bound by $O(nd^k)$ due to marginalization, we incur this cost for each of $r$ bags in the \cjt.   Thus the total cost is $O(rn(dm)^k)$ to calibrate all $\binom{k}{m} = O(m^k)$ pivot queries.

To simplify our analysis, let us also materialize the absorption results (join result of all incoming messages and relations mapped to a given bag) for each bag during calibration (\Cref{sss:msgpassing}).  This does not change the worst-case runtime complexity, and increases the storage cost by at most the size of the base relations.

Notice that these absorption results can be directly used to answer OLAP queries with $k+1$ attributes with no cost in complexity (\Cref{sss:msgreuse}).  Thus, materializing cuboids of up to $k+1$ attributes only requires the cost to calibrate cuboids with $k$ attributes.

More generally, given an OLAP query that groups by $h$ attributes (so it contains $h$ group-by annotations ${\bf A_h}$), it is executed over a \cjt with $k$ attributes by finding the annotations that differ between the pivot and new query ${\bf A_{h-k}}$, and performing message passing over the associated Steiner tree (\Cref{sss:cjtqexec}).  Further, since the calibrated pivot queries span all combinations of $k$ attributes, we simply need to find the pivot query that results in the Steiner tree that spans the fewest bags. 

To summarize, calibration of all pivot queries with $k$ attributes costs $O(rn(dm)^k)$, and cost to execute an OLAP query with $h>k$ attributes is $O(s({\bf A_{h-k}})\times\phi)$, where  $s({\bf A_{h-k}})$ is the number of bags in the Steiner tree spanning ${\bf A_{h-k}}$, and $\phi$ is the size of the absorption result in $O(nd^{h-1})$, which upper bounds the message size.

\subsection{OLAP Construction Procedure}
Suppose we wish to materialize all cuboids with up to $h$ attributes.  Our complexity analysis shows that there is a space-time tradeoff.   To minimize the time complexity, we calibrate pivot queries with $h-1$ attributes, so that materializing cuboids with $h$ attributes is $O(1)$.  However, calibrating pivot queries with fewer attributes reduces build sizes at the expense of larger Steiner trees during cuboid computation.

\subsection{Experiments}

\stitle{Dataset.} 
  Following prior \jt work~\cite{xirogiannopoulos2019memory}, we  created \texttt{synthetic} dataset that contains $r\in[2,8]$ relations with a chain schema: 
$$R(A_1, A_2), R(A_2,A_3),\ldots, R(A_r, A_{r+1}).$$ 
We vary the fanout $f$ between adjacent relations (low=2, mid=5, high=10), and the attribute domain size $d$. For each value of $A_i$ in $R(A_i, A_{i+1})$, we assign $f$ unique values to $A_{i+1}$ with fanout f being implemented by, for each value in $A_i$, assigning f sequential values to $A_{i+1}$, such that the $n^{th}$ value is $n\%d$. Thus, the fanout $f$ is in both directions.  We vary the fanout (and domain) and keep the total join size $d\times f^8$ fixed to be $10^9$. The domain sizes $d$ for different fanouts are $d_{low}=3906250$, $d_{mid}=2560$, and $d_{high}=10$.

\begin{figure}
  \centering
  \includegraphics[width=0.65\columnwidth]{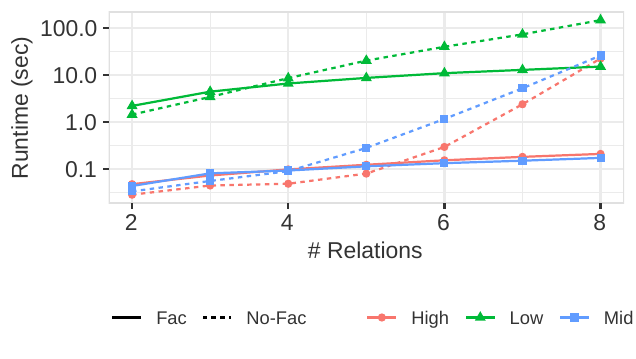}
  \vspace*{-3mm}
  \caption{Run time of total count query with (\texttt{JT})/without message passing (\texttt{No-JT}) in seconds (log scale). High, Mid, and Low are for different fanouts.}
  \label{fig:exp_msg} 
\end{figure}

\subsubsection{Message Passing Costs}
We first evaluate the benefits of message passing (but not calibration) in cloud settings.  The compiler generates \texttt{CREATE VIEW} statements, so that messages are {\it not} materialized.  We execute the total count query as a large join-aggregation query (\texttt{No-JT}) or as an upward message passing (\texttt{JT}).

\Cref{fig:exp_msg} varies the number of relations (x-axis) and fanout (line marker).  Message passing reduces the runtimes from exponential to linear due to early marginalization, but incurs a small overhead to perform marginalization when there are few relations.  Low fan-out has the largest runtime because we fix the total join size and hence the low fan-out has the largest domain size. Note that the x-axis is also interpretable as the Steiner tree size.

For message passing, we can also interpret number of relations as the Steiner tree size, and the run time grows linearly in the Steiner tree size when the bag size is constant.

\begin{figure}[h]
  \includegraphics[width=0.6\columnwidth]{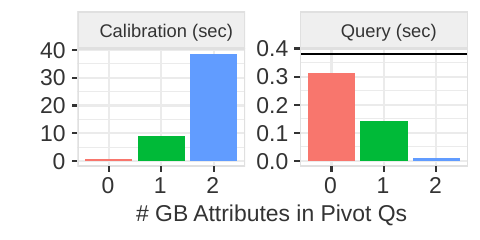}
  \vspace*{-5mm}
  \caption{We vary the dimensionality of the calibrated pivot queries $k \in \{0,1,2\}$ and measure calibration runtime and impact on 4-attribute OLAP queries.  Horizontal line represents the average runtime with \jt.
  }  
  \label{exp:cube_calibration}
\end{figure}

\subsubsection{Cubes in the Cloud} 
\cjts help developers build data cubes to explicitly trade-off build costs and query performance.  To evaluate this, we use the synthetic dataset with $f=10$ (high) fanout and $r=8$ relations, and calibrate all cuboids with $k\in[1,3]$ grouping attributes.  For each $k$, we use the cuboids to execute 100 random OLAP queries with 4 grouping attributes.

The results are in \Cref{exp:cube_calibration}.  Although calibration cost increases exponentially (as expected), message passing is still significantly faster than naive query execution: computing {\it all} 2-attribute cuboids (through calibration of all 1 group-by attribute Pivot Qs in $8.8$s) is substantially faster than naively computing a {\it single} 0-attribute cuboid  ($22.6$s for \texttt{No-JT} in \Cref{fig:exp_msg}). At the same time, increasing the dimensionality of the cuboids ($k$) significantly reduces the query runtimes ($2.71\times$ speedup for $k=1$, and $33.73\times$ speedup for $k=2$) due to the smaller Steiner tree.

The total Redshift table sizes created during calibration is exponential in $k$, and consistent with the analysis in \Cref{app:cube_complexity}.  Redshift appears to pad small tables to ${>}6MB$, hence the large sizes (${\sim}4 GB$ for $k=2$). Unfortunately. the tables cannot be naively compacted (by unioning into a single table) because their schemas are different. Thus we report the actual {\it Data Size} by adding tuple size times cardinality across the tables. The overhead is only $0.17\times$ for $k=0$, $5\times$ for $k=1$ and $127.73\times$ for $k=2$. Given the significant query performance improvement, the space-time trade-off may be worthwhile.

\stitle{Takeaways.} For \cjt, the calibration time grows exponentially in the number of group-by attributes, as there are exponential more pivot queries to calibrate, each takes longer time for larger messages. \cjt is much more efficient than data cube: \cjt can compute all data cuboids up to two group-by attributes in 8.8s, while data cube with naive join takes 22.6s just to compute apex cuboid (\Cref{fig:exp_msg}). The larger number of group-by attributes improves the future OLAP queries performance significantly by reducing Steiner tree size. The reduction of Steiner tree size is especially significant for pivot queries of 2 group-by attributes: given 4 random attributes, there is a high chance ($\sim90\%$) that two of them are in the same relation so only a scan over absorption is needed.

\begin{figure}
  \includegraphics[width=0.8\columnwidth]{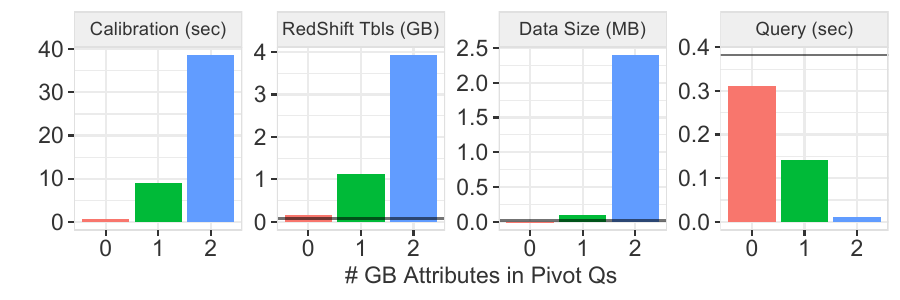}
  \caption{
   RedShift pads tables to be at least $6MB$, which penalizes many small tables.   So we report the total RedShift table size, and actual data size.   Horizontal lines represent, from left-to-right: base DB table size, base DB actual data size, and average runtime with \jt
  }  
\end{figure}

The total Redshift table sizes created during calibration is exponential in $k$, and consistent with the analysis in \Cref{app:cube_complexity}.  Redshift appears to pad small tables to ${>}6MB$, hence the large sizes (${\sim}4 GB$ for $k=2$). Unfortunately. the tables cannot be naively compacted (by unioning into a single table) because their schemas are different. Thus we report the actual {\it Data Size} by adding tuple size times cardinality across the tables. The overhead is only $0.17\times$ for $k=0$, $5\times$ for $k=1$ and $127.73\times$ for $k=2$. Given the significant query performance improvement, the space-time trade-off may be worthwhile.

\section{TPC-H details}
\label{s:tpchqueries}
We discussed how we rewrite TPC-H queries into semi-ring SPJA queries.

\stitle{Query 3.} We remove top and order-by. We also remove $L_ORDERKEY$ group-by because otherwise the result has too many groups.

\begin{verbatim}
SELECT	SUM(L_EXTENDEDPRICE*(1-L_DISCOUNT))	AS REVENUE,
O_ORDERDATE, O_SHIPPRIORITY
FROM	CUSTOMER, ORDERS, LINEITEM
WHERE	C_MKTSEGMENT	= 'FURNITURE' AND
C_CUSTKEY	= O_CUSTKEY AND L_ORDERKEY	= O_ORDERKEY AND
O_ORDERDATE	< '1995-03-28' AND L_SHIPDATE	> '1995-03-28'
GROUP	BY	O_ORDERDATE, O_SHIPPRIORITY;
\end{verbatim}

\stitle{Query 4.}
We rewrite the nested query and remove order-by and distinct.

\begin{verbatim}
SELECT	O_ORDERPRIORITY, count(distinct  O_ORDERKEY)
FROM	LINEITEM, ORDERS
WHERE	O_ORDERDATE	>= '1997-04-01' AND
	O_ORDERDATE	< cast (date '1997-04-01' + interval '3 months' as date) 
AND L_ORDERKEY	= O_ORDERKEY AND L_COMMITDATE	< L_RECEIPTDATE
GROUP	BY	O_ORDERPRIORITY;
\end{verbatim}

\stitle{Query 5.}
For query 5, we break cycle with additional optimization.

\begin{verbatim}
SELECT	N_NATIONKEY,
	SUM(L_EXTENDEDPRICE*(1-L_DISCOUNT))	AS REVENUE
FROM	CUSTOMER, ORDERS, LINEITEM, SUPPLIER, NATION, REGION
WHERE	C_CUSTKEY	= O_CUSTKEY AND	L_ORDERKEY	= O_ORDERKEY AND
	L_SUPPKEY	= S_SUPPKEY AND	C_NATIONKEY	= S_NATIONKEY AND
	S_NATIONKEY	= N_NATIONKEY AND	N_REGIONKEY	= R_REGIONKEY AND
	R_NAME		= 'MIDDLE EAST' AND  o_orderdate >= date '1994-01-01' AND 
o_orderdate < cast (date '1994-01-01' + interval '1 year' as date)
GROUP	BY	N_NATIONKEY;
\end{verbatim}

\stitle{Break cycle for $Q_5$.} $Q_5$ joins customer and supplier by nation, which makes the join graph cyclic and \jt expensive. Luckily, $Q_5$ also group-by nation. We discuss the technique to break cycle: rewrite join + group-by as a set of selections.

Consider the cyclic join of R(A,B), S(A,C), T(B,C). If we know that all future queries will group-by attribute A, we can break the cycle through query rewriting. The original join query is:

\indent\texttt{SELECT A, COUNT(*)}\\
\indent\texttt{FROM R(A,B), S(A,C), T(B,C)}\\
\indent\texttt{WHERE R.A = S.A AND R.B = T.B AND S.C = T.C}\\
\indent\texttt{GROUP BY R.A}

The group by query could be considered as a set of smaller queries, each select a value of A in its domain dom(A). Therefore, for each $a\in dom(A)$, we query

\indent\texttt{SELECT COUNT(*)}\\
\indent\texttt{FROM R(A,B), S(A,C), T(B,C)}\\
\indent\texttt{WHERE \text{\sout{R.A = S.A AND}} R.B = T.B AND S.C = T.C \textcolor{blue}{AND R.A = a AND S.A = a}}

The rewritten query has acyclic join graph. This optimization is closely related to conditioning in Probabilistic graphical model~\cite{koller2009probabilistic}.



\stitle{Query 8.}
We only consider the inner query, as outer query is cheap to compute.

\begin{verbatim}
SELECT	extract(year from o_orderdate) as o_year, 
SUM(L_EXTENDEDPRICE * (1-L_DISCOUNT)), N2.N_NATIONKEY
FROM	PART, SUPPLIER, LINEITEM, ORDERS, 
CUSTOMER, NATION N1, NATION N2, REGION
WHERE	P_PARTKEY	= L_PARTKEY AND S_SUPPKEY	= L_SUPPKEY AND
		L_ORDERKEY	= O_ORDERKEY AND O_CUSTKEY	= C_CUSTKEY AND
		C_NATIONKEY	= N1.N_NATIONKEY AND N1.N_REGIONKEY	= R_REGIONKEY AND 
		R_NAME		= 'ASIA' AND S_NATIONKEY	= N2.N_NATIONKEY AND
		O_ORDERDATE	BETWEEN '1995-01-01' AND '1996-12-31' AND
		P_TYPE		= 'MEDIUM ANODIZED COPPER'
GROUP BY N2.N_NATIONKEY, o_year;
\end{verbatim}

\stitle{Query 9.}
We remove the order-by.

\begin{verbatim}
SELECT	N_NAMEAS NATION, 
extract(year from o_orderdate) as o_year, 
SUM(L_EXTENDEDPRICE*(1-L_DISCOUNT)-PS_SUPPLYCOST*L_QUANTITY)
FROM PART, SUPPLIER, LINEITEM, PARTSUPP, ORDERS, NATION
WHERE S_SUPPKEY	= L_SUPPKEY AND PS_SUPPKEY	= L_SUPPKEY 
AND PS_PARTKEY	= L_PARTKEY AND P_PARTKEY	= L_PARTKEY 
AND O_ORDERKEY	= L_ORDERKEY AND S_NATIONKEY = N_NATIONKEY 
AND P_NAME	LIKE '%green%'
GROUP BY	NATION, O_YEAR
\end{verbatim}

\stitle{Query 10.}
We remove the limit, order-by and modify the interval to 1 day because otherwise, the result size is too large to return without materialization.

\begin{verbatim}
SELECT C_CUSTKEY, C_NAME, SUM(L_EXTENDEDPRICE*(1-L_DISCOUNT))	AS REVENUE, C_ACCTBAL, N_NAME, C_ADDRESS, C_PHONE, C_COMMENT
FROM CUSTOMER, ORDERS, LINEITEM, NATION
WHERE C_CUSTKEY	= O_CUSTKEY	 AND L_ORDERKEY	= O_ORDERKEY		AND O_ORDERDATE	>= '1994-01-01' 
AND O_ORDERDATE < cast (date '1994-01-01' + interval '1 days' 
as date) AND L_RETURNFLAG = 'R'	AND C_NATIONKEY	= N_NATIONKEY
GROUP BY C_CUSTKEY, C_NAME, C_ACCTBAL, 
C_PHONE, N_NAME, C_ADDRESS, C_COMMENT
\end{verbatim}

\end{document}